\documentclass[twocolumn]{aastex631}

\newcommand{\acounits}{\ensuremath{\mathrm{M}_\odot~\mathrm{pc}^{-2}~(\mathrm{K~km~s}^{-1})^{-1}}}
\usepackage{placeins}
\begin{document}
\title{The Radial and Vertical Structure of Molecular Gas in the Edge-On Galaxy NGC 4565}


\author[0000-0002-0932-9879]{Grace Krahm}
\affiliation{Department of Astronomy, The Ohio State University,140 West 18th Avenue, Columbus, OH 43210, USA}

\author[0000-0002-2545-1700]{Adam K. Leroy}
\affiliation{Department of Astronomy, The Ohio State University,140 West 18th Avenue, Columbus, OH 43210, USA}
\affiliation{Center for Cosmology and Astroparticle Physics (CCAPP), 191 West Woodruff Avenue, Columbus, OH 43210, USA}

\author[0000-0003-0378-4667]{Jiayi Sun}
\affiliation{Department of Physics and Astronomy, University of Kentucky, 506 Library Drive, Lexington, KY 40506, USA}

\author[0000-0002-3426-5854]{Kijeong Yim}
\affiliation{Department of Astronomy and Space Science, Chungnam National University, Daejeon 34134, Republic of Korea}

\author[0000-0001-9605-780X]{Eric W. Koch}
\affiliation{National Radio Astronomy Observatory, 520 Edgemont Road, Charlottesville, VA 22903, USA}

\author[0000-0002-7759-0585]{Tony Wong}
\affiliation{Department of Astronomy, University of Illinois, 1002 West Green Street, Urbana, IL 61801, USA}

\author[0000-0003-0645-5260]{Deanne Fisher}
\affiliation{Centre for Astrophysics and Supercomputing, Swinburne University of Technology, Hawthorn, VIC 3122, Australia}
\affiliation{ARC Centre of Excellence for All Sky Astrophysics in 3 Dimensions (ASTRO 3D), Australia}

\author[0000-0002-5204-2259]{Erik Rosolowsky}
\affiliation{Department of Physics, University of Alberta, 4-183 CCIS, Edmonton, Alberta, T6G 2E1, Canada}

\author[0000-0002-4378-8534]{Karin Sandstrom}
\affiliation{Department of Astronomy \& Astrophysics, University of California, San Diego, 9500 Gilman Drive, La Jolla, CA 92093, USA}

\author{Dyas Utomo}
\affiliation{Department of Astronomy, The Ohio State University,140 West 18th Avenue, Columbus, OH 43210, USA}

\author[0000-0003-2552-0021]{Jesse van de Sande}
\affiliation{School of Physics, University of New South Wales, Sydney, NSW 2052, Australia}

\author[0000-0001-5454-1492]{Marie Martig}
\affiliation{Astrophysics Research Institute, Liverpool John Moores University, 146 Brownlow Hill, Liverpool L3 5RF, UK}

\author[0000-0001-9557-5648]{Amelia Fraser-McKelvie}
\affiliation{European Southern Observatory, Karl-Schwarzschild-Straße 2, Garching, 85748, Germany}
\affiliation{ARC Centre of Excellence for All Sky Astrophysics in 3 Dimensions (ASTRO 3D), Australia}

\author[0000-0001-7294-9766]{Michael R. Hayden}
\affiliation{Homer L. Dodge Department of Physics and Astronomy, University of Oklahoma, 440 W. Brooks Street, Norman, OK 73019, USA}

\begin{abstract}
We present high-resolution ($0.94" \approx 55$~pc) ALMA CO(2-1) and $^{13}$CO(2-1) observations of the highly inclined ($i\sim87.5^\circ$) galaxy NGC 4565 covering out to galactocentric radius $\rm R_{gal} \gtrsim \pm 17 \: kpc$. The combination of sensitivity and resolution enables the detection of CO emission well into the \ion{H}{1}-dominated outer disk while isolating individual molecular clouds across the full extent of the galaxy. Although often described as an edge-on Milky Way analog, the molecular gas profile of NGC~4565 has a central gap which is more similar to M31. The $^{13}$CO/$^{12}$CO ratio remains consistent at $0.086\pm0.009$ from $\rm R_{gal}=5-13$ kpc. Based on fits to position-velocity slices, the molecular disk remains thin, with a FWHM scale height of 79.1$\pm$1.6~pc measured from the vertical intensity profile with little evidence for vertical flaring. Molecular clouds in NGC 4565 show sizes, linewidths, and surface densities consistent with those found in similar environments in PHANGS–ALMA galaxies and in M31. We identify a prominent star-forming complex on the ring--an overdensity of molecular gas we term the East Ring Pileup. This feature hosts a compact, multiwavelength-bright region, which we call the Jewel. Effects of galaxy inclination on molecular cloud radius, velocity dispersion, surface density, and virial parameter appear as second-order effects that are strongest in velocity dispersion. At this resolution, GMCs are preferentially aligned with the disk of the galaxy and horizontally elongated by a factor of $\sim2$.

\end{abstract}

\section{Introduction}

\begin{figure*}[ht!]
    \centering
    \includegraphics[width=0.85\textwidth]{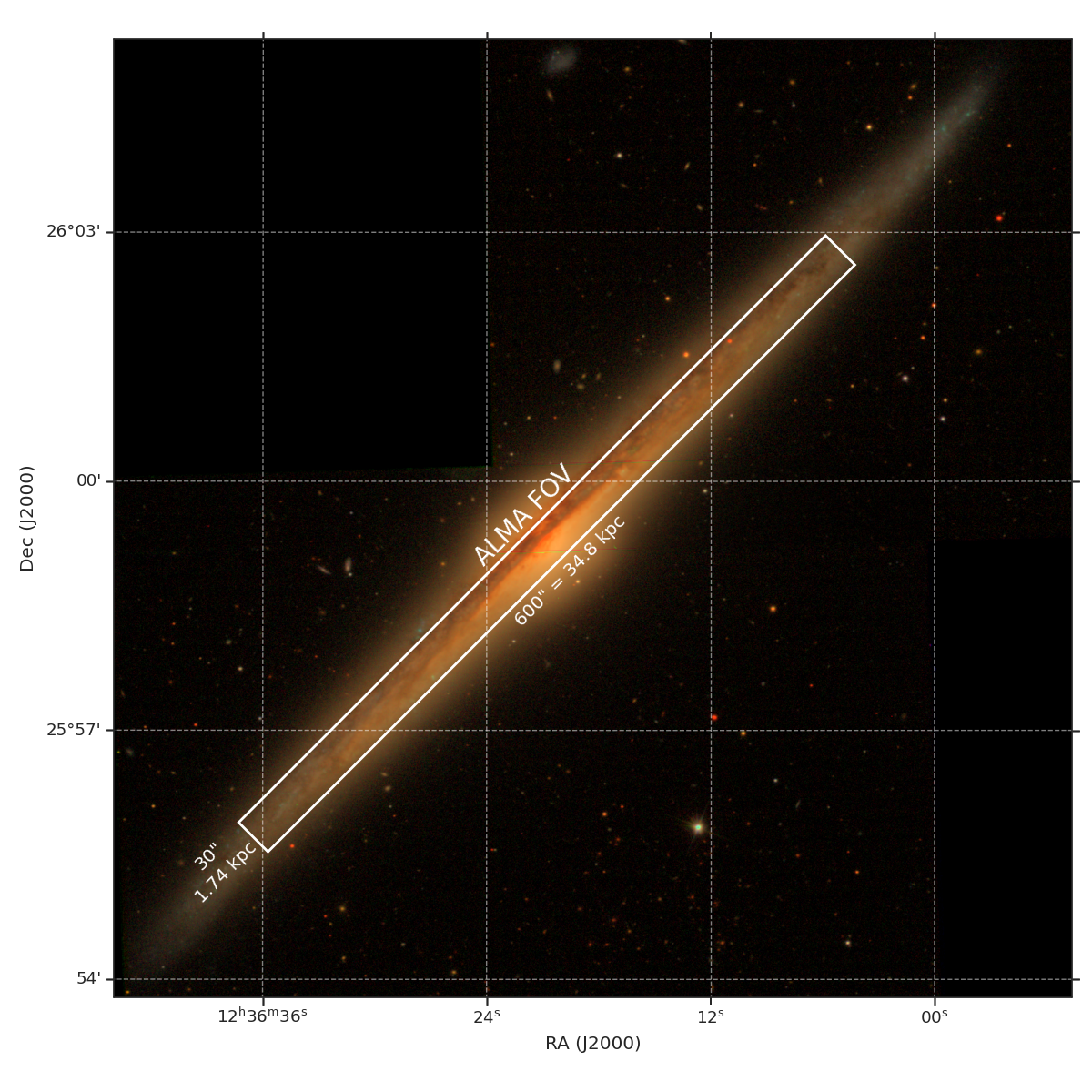}
    \caption{Sloan Digital Sky Survey DR14 \citep{2018ApJS..235...42A} three color image ($g$, $r$, and $i$ bands) with field of view of our new ALMA CO(2-1) survey shown as a white box. Our survey covers most of the stellar disk and is shown is detail in Figure \ref{fig:zooms}.}
    \label{fig:sdss_alma}
\end{figure*}

Understanding the structure and dynamics of the interstellar medium (ISM) is essential for learning about the processes that govern star formation, galaxy evolution, and the cycling of gas between different phases of the ISM \citep{2007ARA&A..45..565M,2017stfo.book.....K}. Star formation occurs in the cold, dense molecular gas, which is organized into giant molecular clouds (GMCs). The physical properties of GMCs-- including their mass, size, and internal motions-- hold information about the conditions under which stars form. The molecular gas volume density ($\rho_{\rm mol}$) is particularly important, as it determines the gravitational free-fall time and is critical to the internal pressure and cloud dynamical state. Through these, it is expected to influence the star formation efficiency \citep{1989ApJ...338..178E,1995ApJS..100..132N,article2, 2005ApJ...630..250K, 2012ApJ...761..156F}. 

Despite its importance, it is difficult to measure $\mathrm{\rho_{\rm mol}}$ directly because GMCs are almost always observed as two dimensional projections (with velocity information) of three dimensional objects. As a result, most extragalactic studies measure molecular gas surface density, $\Sigma_{\rm mol}$, and assume some cloud thickness, $h_{\rm mol}$, along the direction into the sky. This relates to the volumetric density by $\rho_{\rm mol}=\Sigma_{\rm mol}/\sqrt{2\pi}h_{\rm mol}$ (assuming a Gaussian profile), and regions with the same $\Sigma_{\rm mol}$ may have very different $\rho_{\rm mol}$ depending on $h_{\rm mol}$.

Most multiwavelength studies focus on face-on or moderately inclined galaxies to reduce confusion, in which case $h_{\rm mol}$ corresponds to the vertical scale height of the molecular gas layer. One way to constrain the vertical structure of molecular gas is to observe highly inclined or edge-on galaxies. In this case, the vertical thickness of the disk and individual clouds are projected into the plane of the sky. 
However, such observations are challenging: resolving a thin molecular disk, much less individual GMCs, through a crowded and overlapping line of sight requires high angular resolution, high sensitivity, and complex analysis techniques. As a result, only a limited number of studies have targeted molecular gas in highly inclined disks, with the studies of  \citet{2011AJ....141...48Y,2014AJ....148..127Y,2020MNRAS.494.4558Y,2022ApJ...940..118Y} the largest synthetic work to date. However, these pioneering studies still relied mostly on CARMA data with beam sizes larger than the expected $h_{\rm mol}$ and sensitivity to only very massive individual GMCs or collections of GMCs.

In this paper, we present high-resolution ALMA observations of $^{12}$CO(2-1) and $^{13}$CO(2-1) that provide a detailed, high-inclination view of molecular gas in NGC 4565 (Figure \ref{fig:sdss_alma}). This nearby, massive, edge-on spiral galaxy has been considered a Milky Way analog \citep{2019ApJ...872..106K}, though we argue that it also shares many similarities to M31. It has been a classic target for studies of highly inclined galaxies, and is in the sample of \citet{2014AJ....148..127Y}. ALMA's combination of angular resolution ($\approx 0.94" = 55$~pc), sensitivity, and wide spatial coverage ($\sim$35 kpc along the disk) enables us to resolve GMC-scale structures across the entire molecular disk, including in low-density outer regions where atomic gas makes up the majority of the ISM. This extended radial view, coupled with the highly inclined geometry, offers a unique opportunity to examine how molecular cloud properties and the vertical structure of the ISM vary with galactocentric radius and local environment.

In Section \ref{sec:obs}, we describe our new ALMA observations. Section \ref{sec:pano} presents a panoramic view of the molecular disk and models the radial structure of different components of the galactic disk. In Section \ref{sec:gmcs}, we apply a multi-Gaussian fitting method to estimate the vertical extent and linewidth of features that likely correspond to individual GMCs. We use these measurements to study the vertical structure of the molecular disk, the properties of GMCs viewed at high inclination, and the link between GMC properties and environment. 
We summarize our findings in Section \ref{sec:conc}. 

\paragraph{Coordinates} Throughout the paper, we refer to the coordinate along the major axis as the $x$-offset, where a positive distance represents east of the minor axis. Similarly, the coordinate along the minor axis is the $y$-offset, where positive values are north. The $z$ coordinate refers to vertical distance from the midplane of the galaxy in the frame of the disk. With these definitions, $z$ is not orthogonal to $y$. Due to the inclination of the galaxy, the $y$ coordinate encompasses both the foreshortened distance from the center $R_{\rm gal}$, as well as the height above the midplane, $z$, which is mostly not foreshortened. Disentangling these two distances are discussed further in Appendix \ref{sec:yoff}. At the 12 Mpc distance to NGC 4565, $1'' = 60$~pc on the sky, foreshortened in the plane of this disk along the minor axis $1'' \approx 1.3$~kpc.

\section{Observations and Archival Data}\label{sec:obs}

We present ALMA Band 6 observations (2018.1.01050.S; P.I.: Dyas Utomo) observed during Cycles 6 and 7. This includes observations of $^{12}$CO(2-1) and $\mathrm{^{13}CO(2-1)}$, which are summarized in Table \ref{tab:allobs}. 


We combined 7m array observations, which have synthesized beam $\approx 7''$, with 12m array observations obtained in the C-4 ($0.5''$) and C-2 ($1.4''$) configurations. The maximum recoverable scale of the combined dataset is $\sim 34"$ ($\sim 2$~kpc), limited by the shortest baselines in the 7m array. Observations with the Total Power array were also requested but not observed. Across all observing blocks, NGC 4565 was observed for 12.77 hours with the ACA, 5.50 hours with the 12m in C-2, and 12.32 hours in C-4.

We used the standard ALMA pipeline \citep{2023PASP..135g4501H} which is implemented in \texttt{CASA} \citep{2022PASP..134k4501C} to reduce and calibrate the data. The calibrated $uv$ data are imaged with the PHANGS-ALMA pipeline \citep{2021ApJS..255...19L} using Briggs weighting with a robust parameter of 0.5. For $^{12}$CO(2-1), this resulted in a cube with round beam with FWHM 0.94" (55 pc). For the fainter $^{13}$CO line, we analyze a cube including only the C2 12m and 7m data, which has a  1.8" (104 pc) beam but improved surface brightness sensitivity (Table \ref{tab:allobs}).

After imaging, our $^{12}$CO(2-1) map covers 600$''\times30''$ (35 kpc $\times$ 1.7 kpc) with rms noise 78 mK in each 2.5~km~s$^{-1}$ channel. This converts to a $3\sigma$ molecular gas mass surface density and $3\sigma$ point mass sensitivity of 3.9 $M_\odot \rm pc^{-2}$ per channel and $1.5\times10^{4}\rm~M_\odot$ assuming $\alpha_\mathrm{CO(2-1)}=6.7\: \rm (K\:km\:s^{-1}\:pc^{-2})^{-1}$ (see \S \ref{sec:conversions}) and a 2.5~km~s$^{-1}$ line width.
In addition to the CO emission lines from ALMA, we analyze archival infrared data from the \textit{Spitzer} Space Telescope at 3.6 $\mu $m from \citet[][, S4G]{2010PASP..122.1397S} and 8 $\mu $m and 24 $\mu m$ from \citet{2012MNRAS.423..197B}. We also used observations of 21-cm HI emission from the Karl G. Jansky Very Large Array \citep{2014AJ....148..127Y} which we masked using lower resolution (31.8"$\times$13.8") data from the Westerbork Synthesis Radio Telescope \citep{2011A&A...526A.118H}, as well as MUSE H$\alpha$ line maps from the GECKOS survey (J. van de Sande et al. in preparation; \citealp{2024IAUS..377...27V}; \citealp{2025A&A...700A.237F}). The noise and resolution for all these observations are shown in Table \ref{tab:allobs}.

\begin{deluxetable}{ccccc}
\tabletypesize{\scriptsize}
\tablecaption{Observations \& Data Parameters\label{tab:allobs}}
\tablehead{\colhead{Line} & \colhead{Beam FWHM} & \colhead{$\mathrm{\sigma_{rms}}$} & \colhead{Channel} & \colhead{Telescope}\\ 
\colhead{} & \colhead{(\arcsec x \arcsec)} & \colhead{} & \colhead{($\mathrm{km\:s^{-1}}$)} } 
\startdata
$\mathrm{^{12}CO(2-1)}$ & 0.94 $\times$ 0.94 &  0.078 K& 2.54 & ALMA \\
$\mathrm{^{13}CO(2-1)}$ & 1.82 $\times$ 1.82 &  0.024 K& 10 & ALMA\\
HI &6.3$\times$5.6&2.73 K&20& VLA\\
3.6 $\mu m$&1.9 $\times$ 1.9&0.61 MJy/sr & ... & Spitzer IRAC\\
24 $\mu m$&6 $\times$ 6&0.45 MJy/sr & ... & Spitzer MIPS\\
H$\alpha$& $\lesssim 1 \times 1$ &0.40 MJy/sr&57&VLT MUSE\\
\enddata
\tablecomments{$\sigma_{\rm rms}$ indicates the typical rms noise per channel.}
\end{deluxetable}

\begin{deluxetable}{ccc}
\tabletypesize{\scriptsize}
\tablecaption{Parameters for NGC 4565\label{tab:ngc4554params}}
\tablehead{\colhead{Parameter} & \colhead{Value} & \colhead{Reference}} 
\startdata
R.A. (J2000)&12 36 20.8&3.6 $\mathrm{\mu}$m peak\\
Dec. (J2000)&25 59 15&3.6 $\mathrm{\mu}$m peak\\
$v_{sys}$&1230 km s$^{-1}$&\citet{2014AJ....148..127Y}\\
Dist.&12 Mpc&\citet{2025ApJ...978...77B}\\
Incl.&87.5$^\circ$&\citet{2012ApJ...760...37Z}\\
P.A.&135$^\circ$&\citet{2014AJ....148..127Y}\\
SFR&0.67 $\mathrm{M_\odot\:yr^{-1}}$&\citet{sfr}\\
$M_*$&$8.0\times10^{10}~M_\odot$&\citet{2017MNRAS.466.1491H}\\ 
$\rm R_e$&9.86 kpc &3.6 $\mu$m half-mass radius\\
$\overline{Z_\odot}$&$0.99\pm0.14$&$M_*-Z$ relation \citep{2017MNRAS.469.2121S} \\
\enddata
\end{deluxetable}

\section{A Panoramic View of an Edge-On Milky Way Analog}\label{sec:pano}

\begin{figure*}[t!]
    \centering
    \includegraphics[trim={7.3cm .38cm 6.3cm .5cm},clip,
    width=.87\textwidth]
    {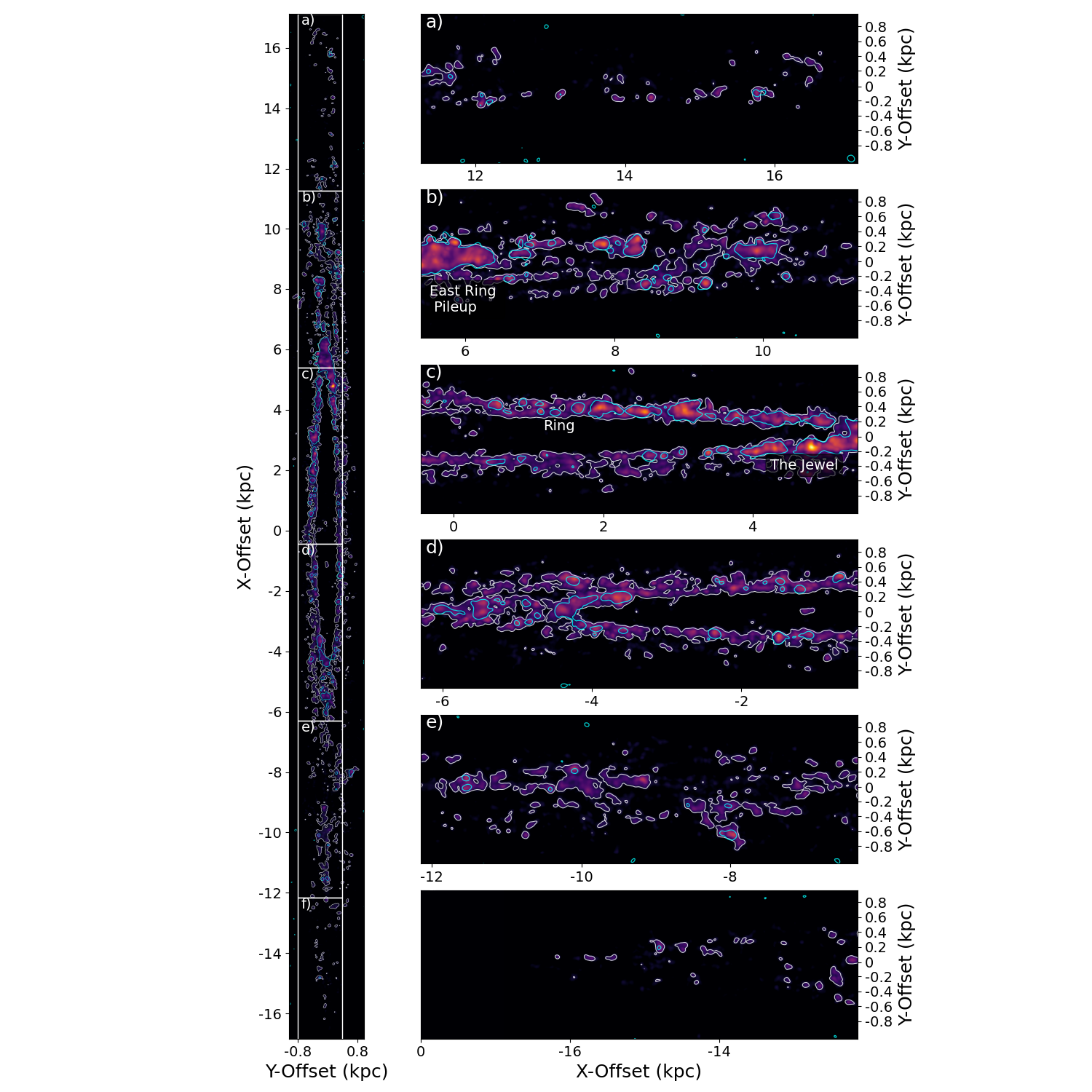}
    \caption{Integrated intensity of $^{12}$CO(2-1) for NGC 4565 on a linear scale [left] with zoom-ins across the galaxy from east to west on an arcsinh scale [right]. Throughout the figures, the $y$-offset is without deprojection. White contours show where $^{12}$CO(2-1) integrated intensity exceeds 5$\sigma$
    and cyan contours indicate where $^{13}$CO(2-1) integrated intensity exceeds 5$\sigma$}.
    \label{fig:zooms}
    
\end{figure*}

\begin{figure*}[t!]
    \centering
    \includegraphics[width=0.49\textwidth]{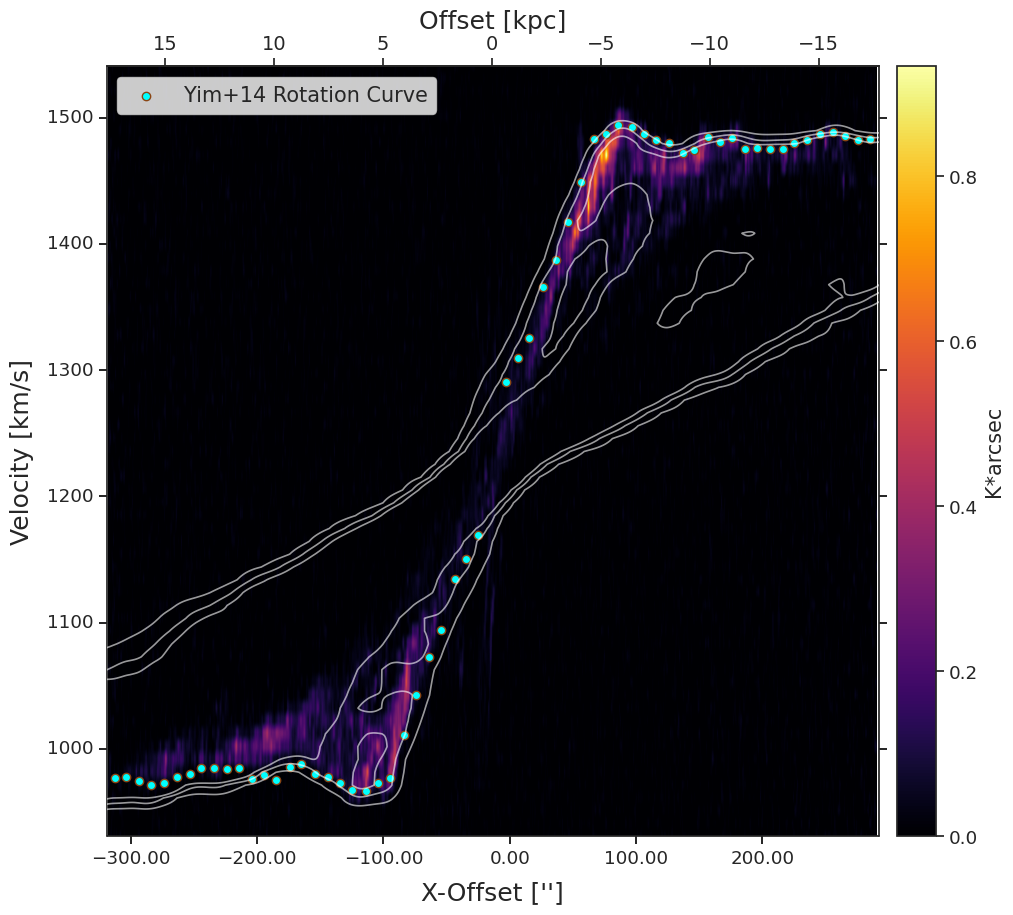} 
    \includegraphics[width=.495\textwidth]{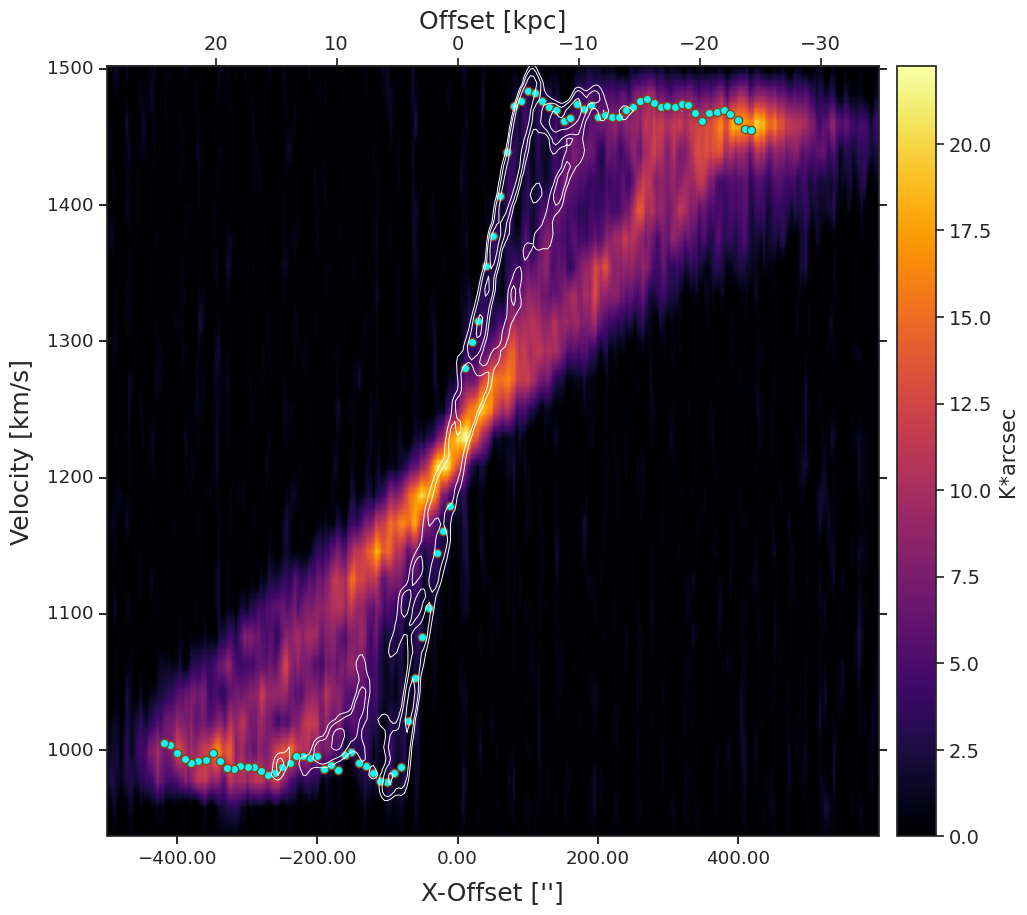} \\
    \includegraphics[trim={0cm 4cm 0cm 4.43cm},clip,width=.99\textwidth]{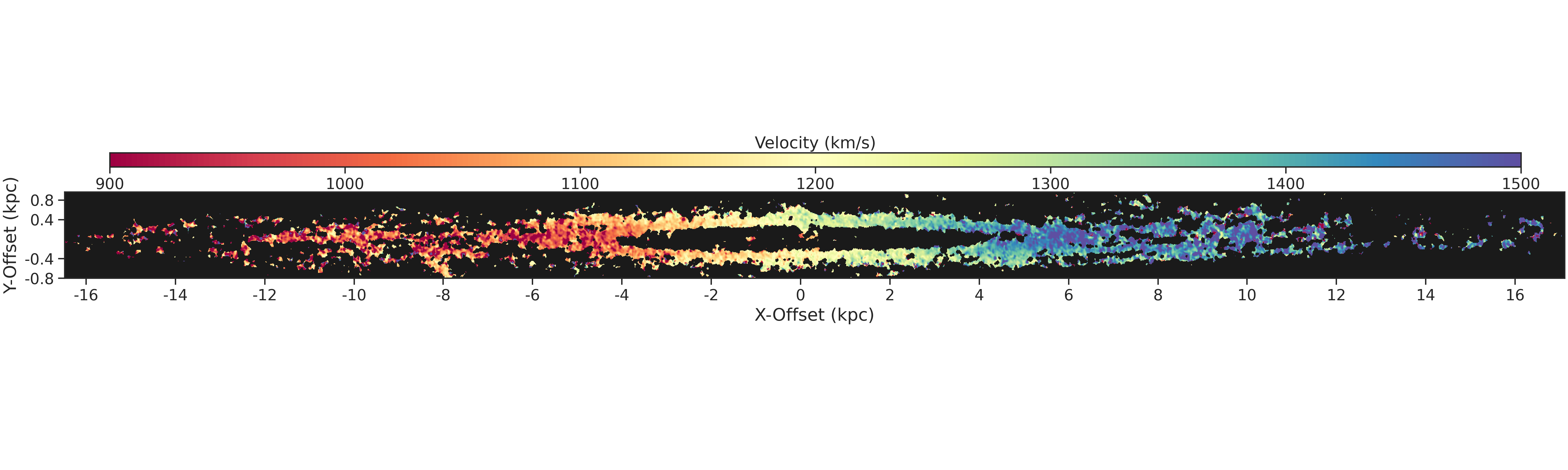}
    \caption{Position–velocity diagrams integrated along the minor axis showing [left] $^{12}$CO(2-1) with HI contours at  5, 10 and 15$\sigma$ for those data and [right] HI with CO contours at 2, 3 and 5$\sigma$ after smoothing to the 6" HI resolution}. The NGC 4565 rotation curve from \citet{2014AJ....148..127Y} is shown in cyan in both panels. The intensity-weighted velocity (moment 1) map of the $^{12}$CO cube is shown in the bottom panel.
    \label{fig:pvd}
\end{figure*}


The ALMA field of view for NGC 4565 is shown in Figure \ref{fig:sdss_alma}. The detailed $^{12}$CO(2-1) integrated intensity map shown in in Figure \ref{fig:zooms} and the position-velocity diagram and velocity field in Figure \ref{fig:pvd}. Because the inclined geometry leads the galaxy to cover a smaller total area on the sky, we are able to map a much larger radial extent compared to most recent ALMA CO mapping of galaxies. We trace emission into the atomic gas–dominated outskirts and over most of the stellar extent of the galaxy.

Because NGC 4565 is slightly less than edge-on, $i = 87.5^\circ$, we are able to distinguish some in-plane features including the inner gaseous ring peaking at $r_{\rm gal} \approx 5$~kpc and individual spiral arms in the outer disk. These structures are clumpy, showing that the molecular gas resolves into individual GMCs at our $\approx$ 55 pc resolution. These GMCs are detected all the way out to the edge of the field of view at $r_{\rm gal} \sim17$ kpc (Figure \ref{fig:zooms}: a. and f.).

Figure \ref{fig:zooms} b. and c. also demonstrate higher surface density in on the eastern side of the ring, which we refer to as the East Ring Pileup. Within this pileup of material, there is a particularly bright region that we label ``The Jewel''. This appears to be a site of recent star formation and also shows concentrated 24 $\mu$m and H$\alpha$ emission (\S \ref{sec:jewel}).

Figure \ref{fig:pvd} shows the gas kinematics for our new $^{12}$CO(2-1) and the VLA HI data. We integrate the CO and HI data cubes along the minor axis, resulting in a position velocity diagram (PVD) along the $x$-axis. Whereas the HI PVD is mostly smooth and symmetric between the east and west sides of the galaxy, the $^{12}$CO appears clumpier and asymmetric, even when smoothed to the coarser resolution of the HI observations (contours on the right panel of Figure \ref{fig:pvd}). Similar to Figure \ref{fig:zooms}, the east side of the ring is brighter than the west, with the extra emission centered around the location of the Jewel at around $50''$ from the minor axis. 

In Figure \ref{fig:pvd}, we plot the rotation curve from \citet{2014AJ....148..127Y}. The galaxy reaches its maximum rotational velocity at the edge of the ring ($x$-offset $\sim100"$) before settling to have approximately flat rotation velocity out to the edge of our field of view. The bottom panel in Figure \ref{fig:pvd} shows the resolved $^{12}$CO(2-1) velocity field.

NGC 4565 has a bar \citep{2019ApJ...872..106K}, and in massive disk galaxies bars often funnel gas inward and concentrate it at dynamical resonances \citep{1996FCPh...17...95B}. However, we do not find clear evidence for gas along bar lanes within the ring or a prominent nuclear gas concentration in Figures \ref{fig:zooms} or \ref{fig:pvd}. These features were also not seen in previous CARMA observations \citep{2014AJ....148..127Y}.

\subsection{Estimating the Galactocentric Radius of Emission}\label{sec:radcomp}

In a nearly edge-on view, each line of sight can intersect a wide range of galactocentric radii. Vertical ($z$) offsets from the galactic midplane can also be confused with in-plane variations in cloud positions along the minor (foreshortened) axis. We therefore combine several estimates of the location of clouds in the disk.

\paragraph{Radius from velocity} Assuming that the radial velocity of the gas is dominated by the circular motions of the galaxy, we employ the PVD method from \citet{2011AJ....141...48Y} in Equations \ref{eqn:rvel} and \ref{eqn:expvalue} to derive a velocity-based galactocentric radius, $R_{\rm vel}$. Assuming circular rotation and a flat rotation curve, for emission at a given major axis offset $x$ and observed radial velocity $V_r$, 

\begin{equation}\label{eqn:rvel}
    R_{\rm vel}=V_c(R_{\rm vel}) \left < \frac{x}{V_r-V_{\rm sys}} \right >,
\end{equation}

\noindent where $V_c$ is the circular velocity from the rotation curve, $V_{sys}=1230\: \rm~km\:~s^{-1}$ is the systemic velocity of the galaxy (Table \ref{tab:ngc4554params}), and the expectation value of \(x/(V_r-V_{sys})\) for emission with velocity width $\Delta V_r=2.5$ km/s (Table \ref{tab:allobs}) can be estimated by

\begin{equation}\label{eqn:expvalue}
    \left < \frac{x}{V_r-V_{\rm sys}} \right > = \frac{|x|}{\Delta V_{r}} \text{ln} \left ( \frac{|V_r-V_{\rm sys}|+\Delta V_r/2}{|V_r-V_{\rm sys}|-\Delta V_r/2}\right )
\end{equation}

\noindent as described in \citet{2011AJ....141...48Y}. 

This method uses velocity information instead of location along the minor axis ($y$), which in principle allows the $y$ location to be leveraged to study the $z$ distribution of material.
However, this approach relies on a well-characterized rotation curve. This is potentially problematic in the inner region of NGC 4565. There is almost no H$_2$ within the 5~kpc ring and HI gas is also scarce and subject to beam smearing in the inner galaxy.

\paragraph{Radius from sky position} We also calculate deprojected galactocentric radii, $R_{dpj}$, assuming a simple thin disk geometry and adopting the disk inclination, position angle, and center in Table \ref{tab:ngc4554params}. For emission without kinematic information like the infrared data, this represents the only available way to estimate the galactocentric radius of individual parcels of emission.

\begin{figure}
    \centering    \includegraphics[width=0.5\textwidth]{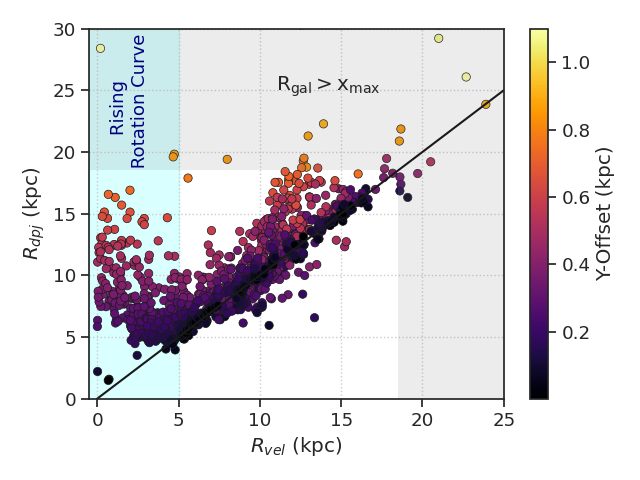}
    \caption{Velocity-based galactocentric radii ($R_{\rm vel}$) compared to radii based on geometric deprojection ($R_{dpj}$) colored by distance along the $y$-axis. The cyan shaded region indicates where the rotation curve is rising (see Figure \ref{fig:pvd}) and the gray shaded region shows where the calculated $R_{gal}$ is larger than the maximum $x$-offset in our map. The black line represents a one-to-one $R_{dpj}=R_{\rm vel}$.}
    \label{fig:radcomp}
\end{figure}

Figure \ref{fig:radcomp} compares the PVD and deprojection methods for CO peaks identified in the next section. The two methods give overall similar results, but differ in the center. This is mainly because
the rotation velocity near the minor axis is difficult to accurately characterize, with most rotational motion lying tangential to the line of sight, i.e., in the plane of the sky. There is also a feature around $R_{\rm vel}\sim12$ kpc in in which $R_{dpj} > R_{\rm vel}$. These clouds show a significant $y$ offset, but their velocity is consistent with only a moderate galactocentric radius. This makes them good candidates to be offset from the mid-plane of the disk in the $z$ direction. In Appendix \ref{sec:yoff}, we explore this using the difference between $R_{\rm vel}$ and $R_{dpj}$ to estimate height above the midplane.

\paragraph{Modeling radial profiles from stripe integrals}
We also model the ensemble of emission in order reconstruct the large-scale radial distribution of material in the disk. We construct multiwavelength radial profiles of the galaxy with the \citet{2025arXiv251005214S} implementation of the stripe integral method \citep{1988A&AS...72...19W}. First we measure stripe integrals by integrating all emission along the minor ($y$) axis in bins defined along the major $x$ axis. Then we invert the set of integrated fluxes to reconstruct what profile of face-on surface brightness is needed to produce the observed stripes. As discussed by \citet{2025arXiv251005214S}, this is formally an inverse Abel transform, though the inversion has also been approached as a deconvolution problem using Lucy-Richardson deconvolution  \citep{1972JOSA...62...55R,1974AJ.....79..745L}. This method does not need velocity information and allows us to reconstruct the large scale structure of the disk using an approach that treats the infrared, HI, and CO data symmetrically. It is not useful for describing the location of a particular region in the galaxy.

\subsection{Surface Densities of Physical Quantities}\label{sec:conversions}

We convert from $^{12}$CO(2-1) and HI 21-cm to surface mass densities using the conversion factors:

\begin{equation}\label{eqn:co}
    \frac{\Sigma_{\rm mol}}{{\rm M}_\odot~{\rm pc}^{-2}} \approx \alpha_{CO}^{2-1}\left( \frac{I_{\mathrm{CO(2-1)}}}{1 \: {\rm K \:km \: s}^{-1}} \right )
\end{equation}

\begin{equation}\label{eqn:hi}
    \frac{\Sigma_{\rm atom}}{{\rm M}_\odot \, \mathrm{pc}^{-2}} = 0.0196  \left( \frac{I_{\rm HI}}{\mathrm{1K\,km\,s}^{-1}} \right)
\end{equation}

\noindent where 
we have assumed a contribution of helium by mass ($\mu_H$) of $\sim$36\% for both phases.  For atomic gas we assume optically thin emission and adopt the conversion factor from  \citet{2011piim.book.....D}. For the CO-to-H$_2$ conversion factor, we use a radially varying $\alpha_{CO}^{2-1}$ following \citet{2025arXiv251005214S}, which estimates the metallicity based on \citet{2017MNRAS.469.2121S} and follows the prescription of \citet{2024ARA&A..62..369S} to relate mass, metallicity, and star formation rate at each radius to an $\alpha_{CO}^{2-1}$ value. This gives a radial profile of $\alpha_{CO}^{2-1}$ which ranges from 2.01 -- 16.59 \acounits\ across the full field of view with an average $\alpha_{CO}^{2-1}$ of 9.45 \acounits. However, most of the gas is in the ring, rather than in the center or far outskirts of the galaxy. Near the ring ($4.68\leq R_{gal}\leq$6.55 kpc), $\overline{\alpha_{CO}^{2-1}}=7.28\pm0.50$ \acounits\ which close to a typical Milky Way-like $\alpha_{CO}^{2-1}=6.7$ \acounits \citep{2013ARA&A..51..207B,2021MNRAS.504.3221D}.





The IR data are converted to surface density profiles using Equations \ref{eqn:star} and \ref{eqn:sfr} following \citet{2019ApJS..244...24L}.

\begin{equation}\label{eqn:star}
    \frac{\Sigma_{*}}{{\rm M}_\odot ~{\rm pc}^{-2}} \approx 350 \left( \frac{\Upsilon_\star}{0.5~{\rm M}_\odot / {\rm L}_\odot} \right) \left( \frac{I_{\mathrm{3.6\:\mu m}}}{1 \: \textrm{MJy~sr}^{-1}} \right ) 
\end{equation}

\begin{equation}\label{eqn:sfr}
    \frac{\Sigma_{\rm SFR}}{{\rm M}_\odot~{\rm Gyr}^{-1}~{\rm pc}^{-2} } \approx 2.97 \left( \frac{I_{\mathrm{24\:\mu m}}}{1 \: \textrm{ MJy~sr}^{-1}} \right ) 
\end{equation}

Based on the radial profile of specific star formation rate (SFR/M$_\star$) we expect the near-infrared stellar mass to light ratio to remain at $\Upsilon_\star \approx 0.5~{\rm M}_\odot / {\rm L}_\odot$ across the galaxy using prescriptions from \citet{2019ApJS..244...24L} as implemented by \citet{2025arXiv251005214S}. Eq. \ref{eqn:sfr} assumes that the mid-infrared emission captures all of the star-forming activity and that we do not need additional ultraviolet or H$\alpha$ observations to estimate $\Sigma_{\rm SFR}$. Given NGC 4565's inclined geometry and thus high optical depth, this should be a reasonable assumption.  The more likely source of bias is including ``diffuse'' mid-infrared emission that reflects dust heated by an older stellar population \citep[e.g.,][]{2023ApJ...944L...9L,Belfiore_2023,2024AJ....167...39P}. Since the stripe integrals reconstruct the face-on surface brightness values, it is not necessary to additionally correct them for inclination. 


\subsection{ISM Composition and Radial Structure}\label{sec:radial_analysis}

In Figure \ref{fig:radprof} we show azimuthally averaged profiles of $^{12}$CO(2-1) and $^{13}$CO(2-1), molecular gas mass surface density, and other disk components. Together, these reveal the large-scale structure of molecular gas in NGC 4565 and the changing phase makeup of the ISM.

\subsubsection{Radial profiles of gas and stars}

As demonstrated in Figure \ref{fig:zooms}, there is very little molecular gas in the inner parts of the galaxy. This is reflected in the radial profiles with molecular gas and star formation rate surface densities, $\Sigma_{\rm mol}$ and $\Sigma_{\mathrm{SFR}}$, exhibiting central depressions. Both quantities peak near the ring ($R_{\rm gal} =4.5$ kpc) and decline approximately exponentially outside this radius. The atomic gas ($\Sigma_{\mathrm{HI}}$) shows low values within the ring in the map and then remains relatively flat and low across the disk. The stellar surface density ($\Sigma_*$) shows a central peak and smooth exponential decline, with a mild enhancement at the ring radius.


To quantify the structure of our derived surface density profiles, we fit exponential disk models:

\begin{equation}\label{eqn:exp}
    \Sigma (r) = \Sigma_{r_0} \exp \left ( - \frac{r-r_0}{l} \right) ,
\end{equation}
where $\Sigma_{r_0}$ is the surface density at $r_0$, $l$ is the scale length, and $r_0$ is the offset from the center. 

The fitted exponential disk models are plotted in Figure \ref{fig:radprof}a with the best-fit parameters in Table \ref{tab:expdisk_combined}. $\Sigma_*$ follows a smooth exponential decline from the galaxy center with a scale length of 4.5 kpc. This is slightly larger than the 4 kpc value reported by \citet{2014AJ....148..127Y} which is likely due to \citet{2014AJ....148..127Y} using a different fitting function that fits for both the scale length and scale height. $\Sigma_{\mathrm{SFR}}$ and $\Sigma_{\rm mol}$ show comparable exponential behavior beyond the ring, with scale lengths of 6.8 kpc. Outside the ring, the atomic gas remains flat at 4 $\rm M_\odot \: pc^{-2}$ to the edge of our $^{12}$CO map. Beyond $R_{\rm gal}\approx 15$~kpc, it begins to drop off exponentially with a scale length of 7.7 kpc. None of the surface density profiles show evidence of a steep drop before the map edges that would signify a cutoff of the disk. 

\subsubsection{ISM phase breakdown and gas fraction}

With $\Sigma_{*}$ decreasing much faster than $\Sigma_{\rm atom}$, the ratio of $\Sigma_{\rm gas}/\Sigma_{*}$ steadily increases with galactocentric radius. At $R_{\rm gal} \sim$~23 kpc, the HI begins to dominate over the stars (Table \ref{tab:expdisk_combined}) and the baryonic mass of the galaxy disk is primarily composed of diffuse neutral gas. 

In the inner galaxy, the molecular gas mass is comparable to the atomic gas mass, $\Sigma_{\rm mol} \gtrsim \Sigma_{\rm atom}$ until $R_{gal}\sim 7$~kpc and then $\Sigma_{\rm atom}$ becomes dominant beyond $R_{\rm gal} \sim 12$ kpc (Figure \ref{fig:radprof} a, d; Table \ref{tab:expdisk_combined}). Consistent with molecular gas being the site of star formation, the radial profile of H$_2$ closely resembles that of $\Sigma_{\rm SFR}$.



Similarly to the $\Sigma_{*}/\Sigma_{\rm HI}$ and $\Sigma_{\rm mol}/\Sigma_{\rm HI}$, the $\Sigma_{\rm SFR}/\Sigma_{\rm gas}$ (Figure \ref{fig:radprof} c) steadily decreases with increasing $R_{gal}$ since the SFR falls much quicker than the total gas content. However, the specific star formation rate (sSFR $\equiv \Sigma_{\rm SFR}/\Sigma_{*}$) maintains a consistent 0.013 Gyr$^{-1}$. Meanwhile, the star formation efficiency of the molecular gas ($\Sigma_{\rm SFE}^{\rm mol}\approx\Sigma_{\rm SFR}/\Sigma_{\rm mol}$) is also relatively constant from 5-15 kpc.

\subsubsection{Comparison to other galaxies}
In Figures \ref{fig:radprof}e and f, we compare the radial profiles of NGC 4565 to those of the Milky Way \citep{2024arXiv240702859S}, M31 \citep{2025ApJS..279...35K, 2006A&A...453..459N}, and kpc-scale radial profiles in 24 star forming galaxies with similar stellar masses (10.5 $\rm M_\odot<$ log $M_*<11.1\: M_\odot$) from various surveys. This includes $^{12}$CO(2-1) and HI data from PHANGS \citep{2021ApJS..257...43L,2024A&A...691A.163E,2025MNRAS.536.2392C}, HERACLES \citep{2009AJ....137.4670L},
VERTICO \citep{2021ApJS..257...21B}, VIVA \citep{2009AJ....138.1741C}, THINGS \citep{2008AJ....136.2563W}, and EveryTHINGS \citep[P.I. K. M. Sandstrom]{2025MNRAS.536.2392C}.


The molecular gas in NGC 4565 shows lower surface densities in the inner galaxy due to the empty ring, but a larger H$_2$ scale length. $^{12}$CO is detected out to $R_{\rm gal} \gtrsim 15$~kpc at surface densities higher than the comparison galaxies. Thus NGC 4565 appears to show an unusually extended molecular gas disk --- we measure $l_{\rm mol} / l_\star \approx 1.5$ across the whole $H_2$ disk and $l_{\rm mol} / l_\star \approx 1.23$ for 5 kpc $\leq R_{\rm gal}\leq$ 15 kpc, while many literature studies find $l_{\rm mol} \approx l_\star$ on average \citep[e.g.,][]{2009AJ....137.4670L,2017ApJ...846..159B,2021ApJS..257...21B}. NGC 4565 also stands out for its relatively low $\Sigma_{\rm HI}$ values compared to the rest of the sample.

We highlight the Milky Way, NGC 2775, and M31 as potential less-inclined analogs. The Milky Way shows a two-peaked molecular gas profile, with the brightest $^{12}$CO emission in the center, a smaller peak at the ring, and less depression within the ring compared to NGC 4565, presumably reflecting molecular gas along the bar. The Milky Way also shows a short CO scale length compared to NGC 4565 or the comparison sample. This might reflect the influence of the bar or the distance assigned to the gas being uncertain with regards to the bar ends. Recent claims that the Milky Way is less massive than previously believed (log $M_*$ = 10.4 $M_\odot$; \citealt{2025ApJ...990L..37L}) could also be consistent with a shorter scale length. The difference may also reflect methodological differences in construction of the profiles. Regardless, NGC 4565 and the Milky Way do not appear to be perfect analogs in their molecular gas structure.



NGC 2775 is the clearest case of an empty ring in the PHANGS-ALMA sample. It shows little molecular gas in the center and has a low maximum $\Sigma_{\rm HI}$, both similar to NGC 4565. However the galaxy lacks a bar and only the ring is clearly detected in CO and HI. NGC 2775 thus lacks the extended disk see in NGC 4565, though sensitivity effects may play some role here. 

M31 may represent the best qualitative analog among out sample. Like NGC 4565, M31 lacks a central molecular gas concentration or strong CO emission along its bar. The resemblance is only qualitative, however. M31's most prominent molecular gas ring lies at $R_{\rm gal} \sim 10$~kpc, almost twice that of NGC 4565's ring. M31 also shows more emission interior to its ring, and a much higher maximum $\Sigma_{\rm HI}$. It also has moderately lower $\Sigma_{\rm mol}$ at intermediate radii than NGC~4565.

Both NGC 4565 and M31 are interacting with companion galaxies. M31's interactions with M33 and especially M32 have influenced its ring structure and warping of the disk \citep{2006Natur.443..832B}. Likewise, NGC 4565 shows a clear large-scale warp that has been attributed to its interaction with the nearby dwarf galaxy IC 3571 \citep{2012ApJ...760...37Z}. The NGC 4565–IC 3571 interaction might also be an important factor in redistributing gas within the disk and creating the galaxy's prominent ring and extended molecular disk. 

\begin{deluxetable}{c|ccc}
\tabletypesize{\scriptsize}
\tablecaption{Exponential Disk Fits\label{tab:expdisk_combined}}
\tablehead{ \colhead{Profile} & \colhead{Scale Length} & \colhead{$\Sigma_0$} & \colhead{$r_0$}\\ \vspace{-1cm} & \colhead{(kpc)\rule{0pt}{10pt}}&\colhead{($M_\odot\:\mathrm{pc}^{-2}$)} &  \colhead{kpc}}
\startdata
$\Sigma_{\rm mol}$      & 6.82$\pm$0.39  & 11.82$\pm$0.84 &4.68\\ 
$\Sigma_{\rm HI}$       & 7.68  $\pm$0.35& 20.50 $\pm$4.2&14.62\\ 
$\Sigma_{\rm *}$        & 4.48$\pm$0.12    & 401.4$\pm$12.6 &2.50\\ 
$\Sigma_{\rm SFR}$\tablenotemark{a} & 6.83  $\pm$0.17& 1.87 $\pm$0.13&3.80\\ 
\enddata
\tablenotetext{a}{The units of $\Sigma_{\rm SFR}$ and $\Sigma_{\rm SFR,0}$ include an additional factor of Gyr$^{-1}$.}

\end{deluxetable}

\subsection{\texorpdfstring{$^{13}$CO--to--$^{12}$CO ratio}{13CO--to--12CO ratio}} 
\label{sec:12to13}

\begin{figure*}
    \centering
    \includegraphics[trim={1.6cm 2.95cm 2.95cm 4.95cm},clip,width=0.9\textwidth]{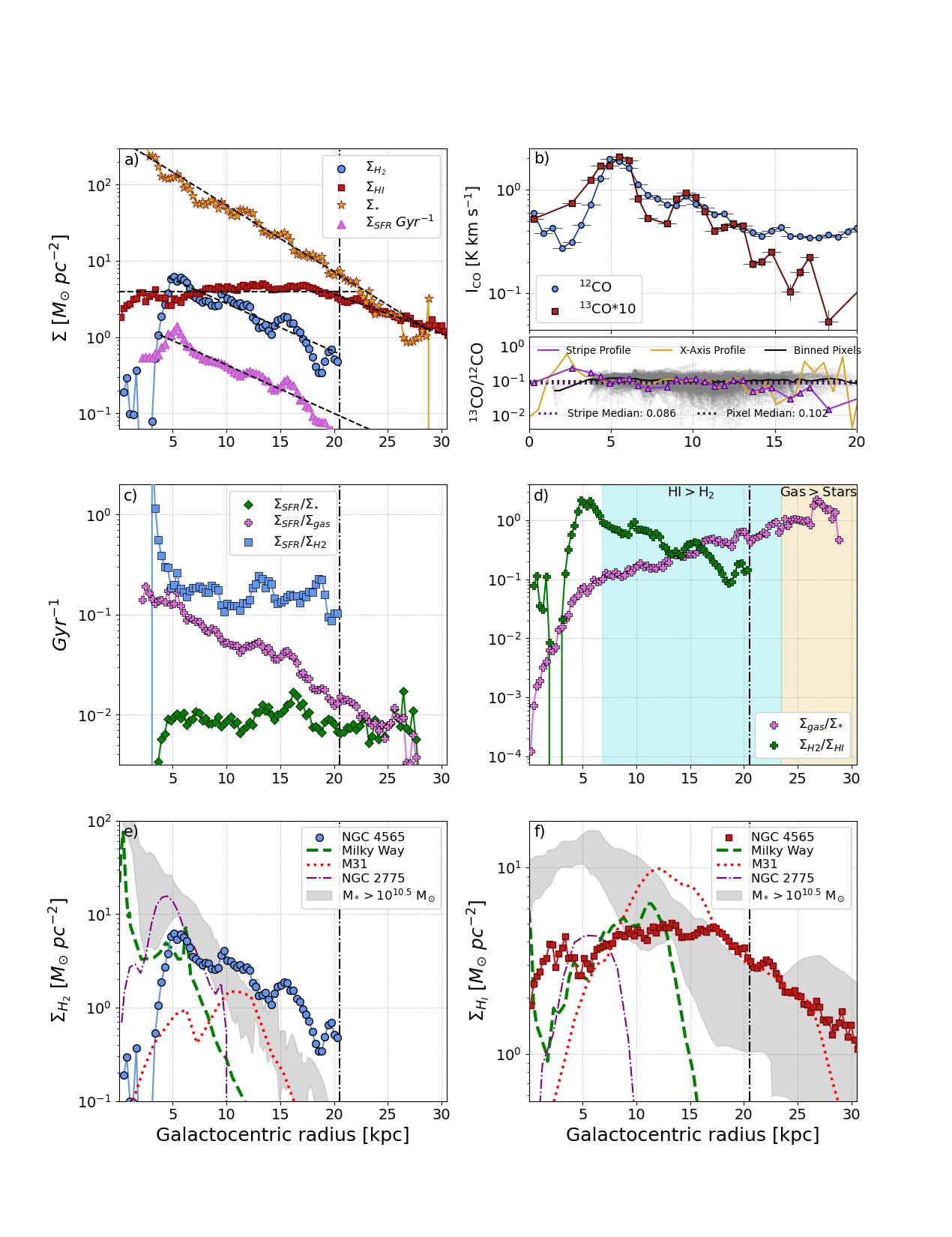}
    \caption{Radial profiles of NGC 4565 with stripe integrals out to 30 kpc. The edge of the $^{12}$CO map is shown as a vertical black line. [a] Stellar, molecular gas (H$_2$), atomic gas (HI), and SFR surface density. The SFR surface density has an extra factor of Gyr$^{-1}$ for full units of $\mathrm{M_\odot \;pc^{-2}\;Gyr^{-1}}$}. Exponential disk fits are shown with dashed black lines. [b] $^{12}$CO(2-1) and $^{13}$CO(2-1) integrated intensity radial profiles from the stripe integral modeling with the $^{13}$CO/$^{12}$CO profile plotted below from stripe profiles, $x$-axis profiles, and pixel-based ratios in radial bins. [c] Star formation rate surface density per stellar, total gas, and H$_2$ surface density. [d] Dimensionless ratios of H$_2$, HI, and stars. [e \& f] H$_2$ and HI profiles of NGC 4565 from compared to the Milky Way, M31, and NGC 2775, as well as the 25-75\% range of similarly massive galaxies in lower resolution surveys.
    \label{fig:radprof}
\end{figure*}

Figure \ref{fig:radprof}b compares $^{12}$CO(2-1) and $^{13}$CO(2-1) as a function of galactocentric radius. Both are detected out to the the edge of the map at $x$-offset $\sim$17.5 kpc. The $^{13}$CO(2-1) cube is masked using the mask that we apply to the $^{12}$CO(2-1) cube, so that the emission used for the profiles are the same. We calculate the $^{13}$CO(2-1)/$^{12}$CO(2-1) ratio in three ways: using the stripe integrals as described above, summing in the emission after binning along the major ($x$) axis, and taking the median among only detected pixels.

The optically thinner $^{13}$CO(2-1) line yields a median $^{13}$CO/$^{12}$CO ratio of $0.086\pm0.01$, with slight differences between the results for radial profile and pixel-by-pixel ratios. This is almost exactly the median $\rm ^{13}CO(1-0)/^{12}CO(1-0)$ ratio found by \citet{2018MNRAS.475.3909C} of $0.089\pm0.004$ for a sample of massive disk galaxies. The ratio reflects a combination of the isotopologue abundance and the optical depth of the lines. For a commonly assumed Solar Neighborhood abundance of $\approx 60$ \citep[][see \citealt{2018MNRAS.475.3909C}]{1994ApJ...432..148W,2005ApJ...634.1126M}, our adopted ratio implies a $^{12}$CO(2-1) optical depth of $\approx 5$, in good agreement with previous work.

Our profile in Figure \ref{fig:radprof}b reaches almost twice the galactocentric radius of any of the \citet{2018MNRAS.475.3909C} profiles or the \citet{2016ApJ...818..144R} profile of the Milky Way. Even out to this large radius, the $^{13}$CO/$^{12}$CO ratio is relatively flat. Both the stripe model and the $x$-axis profile suggest an enhanced ratio in the ring, then all three methods find consistent, nearly fixed values from $R_{\rm gal} \approx 5{-}13$~kpc. Outside that, the stripe and $x$-axis profiles suggest a moderate drop in the ratio, with $^{13}$CO becoming fainter. The pixel ratio remains high because our sensitivity limits prevent detecting individual faint $^{13}$CO pixels. Beyond only radial behavior, the $^{12}$CO profile is smoother than the $^{13}$CO one. This suggests that the ratio may also show opacity variations, e.g., associated with spiral arms or other features beyond just a radial trend.


Though a single gradient is not an ideal description, the stripe integral profiles suggest an overall decline of $-0.036\text{ dex kpc}^{-1}$, from $^{13}$CO/$^{12}$CO=0.18 near the ring to 0.03 beyond $R_{\rm gal} \approx 15$~kpc. In the Milky Way, the $^{12}$C/$^{13}$C ratio increases with galactocentric radius \citep{1990ApJ...357..477L, 2024A&A...690A.372L}. The higher values in the ring and lower values in the outer disk may reflect a similar trend in NGC 4565. The flat profile over a large range of intermediate radii might imply increasing optical depth with increasing $R_{\rm gal}$ to offset a dropping $^{12}$C/$^{13}$C ratio.




\section{Physical Properties of Highly Inclined GMCs}\label{sec:gmcs}


\subsection{Methods}\label{sec:methods2}

Methods of determining the scale height of gas in galaxies tend to fall into three main categories: axisymmetric, double Gaussian, and cloud-based calculations \citep{2015ARA&A..53..583H}. The axisymmetric method assumes a smooth molecular disk and was used by several early studies of the Milky Way molecular disk \citep{1975ApJ...202...30B,1975ApJ...199L.105S,1988ApJ...324..248B}. This approach describes the relationship between the density profile, scale height, and offset from the midplane using a single exponential.

The full vertical profile of a galaxy is not always well-described by a single Gaussian (see \citealp{1985ApJ...297..751D}). In the case where both the near and far side of the galaxy are visible, there should be at least two Gaussian components along each line of sight,  one for the near side and one for the far side. In highly inclined but not 90$^\circ$ disks, the components from the near and far sides can also appear at distinct positions in addition to different velocities. Accounting for these two components is referred to as the double Gaussian method and can be used to interpret both Galactic \citep{1988ApJ...327..139C,2006PASJ...58..847N} and extragalactic \citep{2014AJ....148..127Y} observations.

Both the axisymmetric and double Gaussian methods assume a smooth disk, and thus work best on atomic or low-resolution molecular gas observations. As higher resolution observations start to show the clumpy nature of H$_2$. Then the assumption of smoothness breaks and a cloud-based approach is needed. Some Galactic studies have adopted cloud-based approaches, using \textsc{Hii} regions \citep{1990A&A...230...21W} or GMCs identified with algorithms like \texttt{CLUMPFIND} \citep{1994ApJ...428..693W} or \texttt{CPROPS} \citep{2006PASP..118..590R}. This has the advantage of not requiring the disk to be perfectly edge on or smooth. However, these approaches only reveal the vertical structure of each individual feature unless the three dimensional location of each cloud can be determined.

\begin{figure*}[h]
    \centering
    \includegraphics[width=0.9\textwidth]{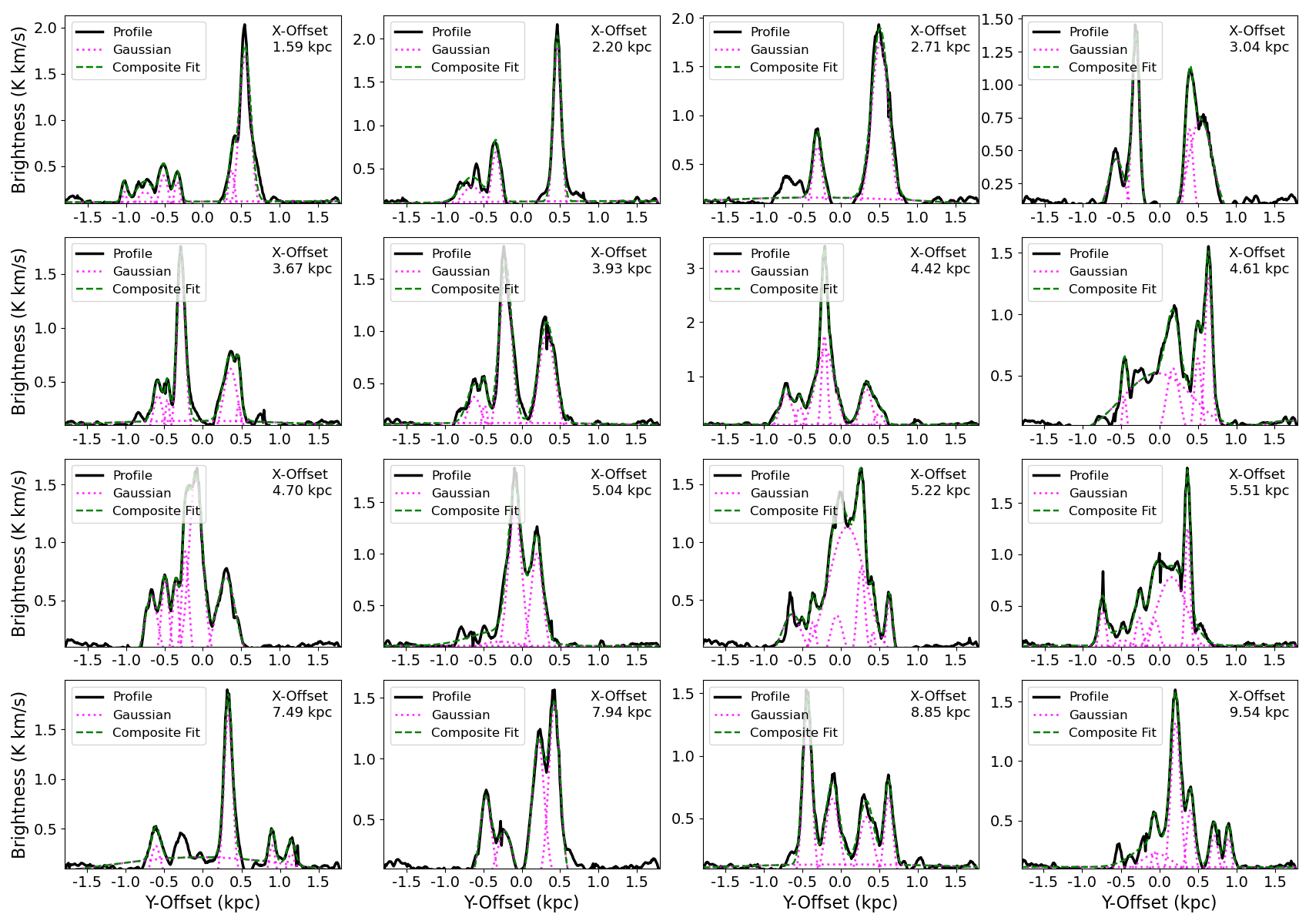}
    \caption{Vertical ($y$-offset) profiles in 16 bins along the major ($x$) axis of NGC 4565. The bin size is 2 beams ($\sim2\arcsec$) wide. The individual fitted Gaussian components are shown by the dotted magenta lines and the green dashed lines show the composite fit of all the Gaussians.}
    \label{fig:vertprof}
\end{figure*}

\paragraph{Gaussian fitting in position-velocity space} Figure \ref{fig:zooms} shows that our high resolution, combined with the less-than-edge-on $i=87.5^\circ$ orientation of NGC 4565, causes the CO emission to resolve into many discrete features. 
Figure \ref{fig:vertprof} shows the $y$-axis intensity profiles for a portion of the galaxy. Each shows many individual components. These features would be labeled as individual molecular clouds in studies of more face-on galaxies at similar resolution.

\begin{figure*}
    \centering
    \includegraphics[trim={.3cm 4.5cm 0cm 4.5cm},clip,width=0.49\textwidth]{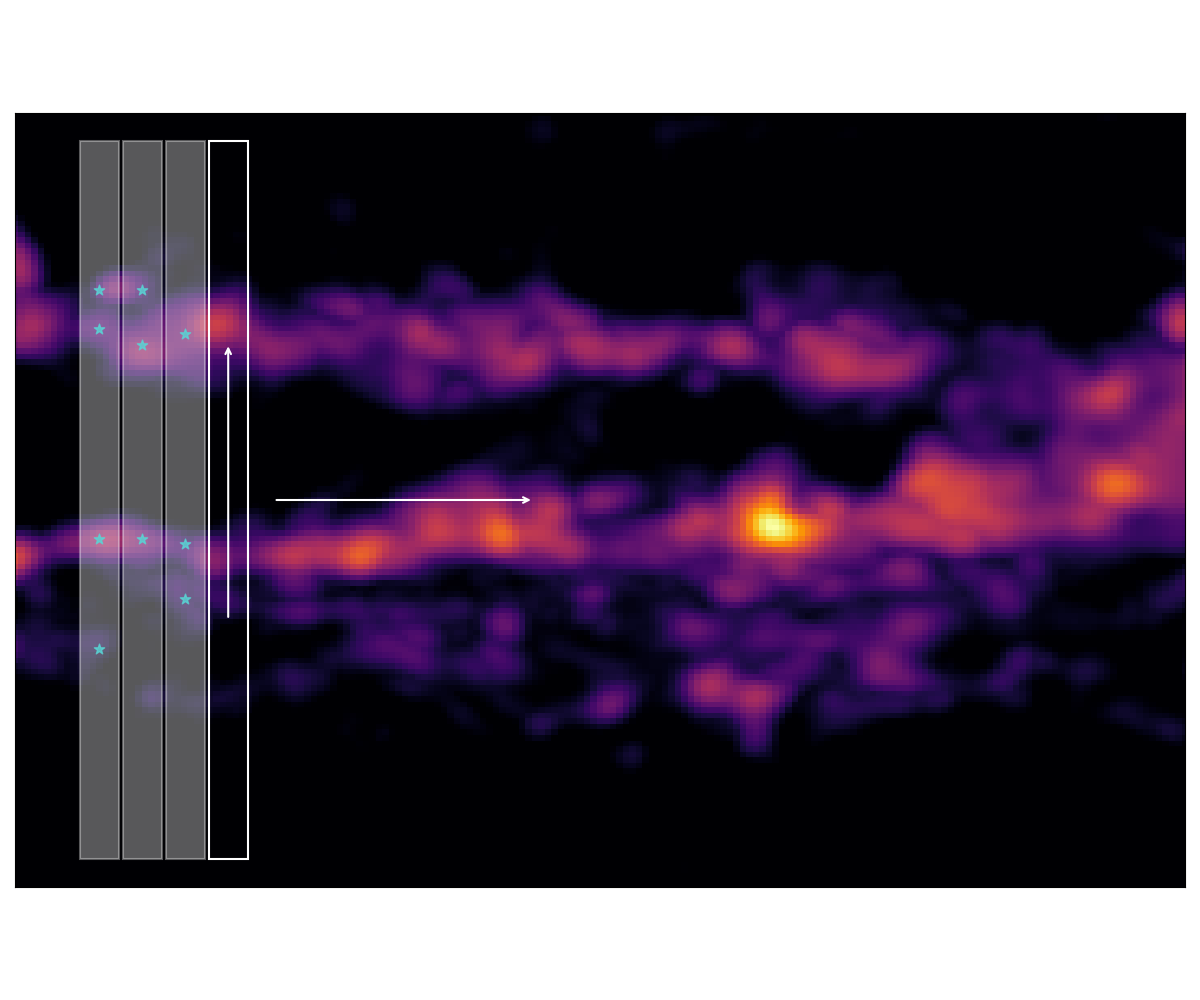}
    \includegraphics[trim={.94cm .6cm .35cm .45cm},clip,width=0.49\textwidth]{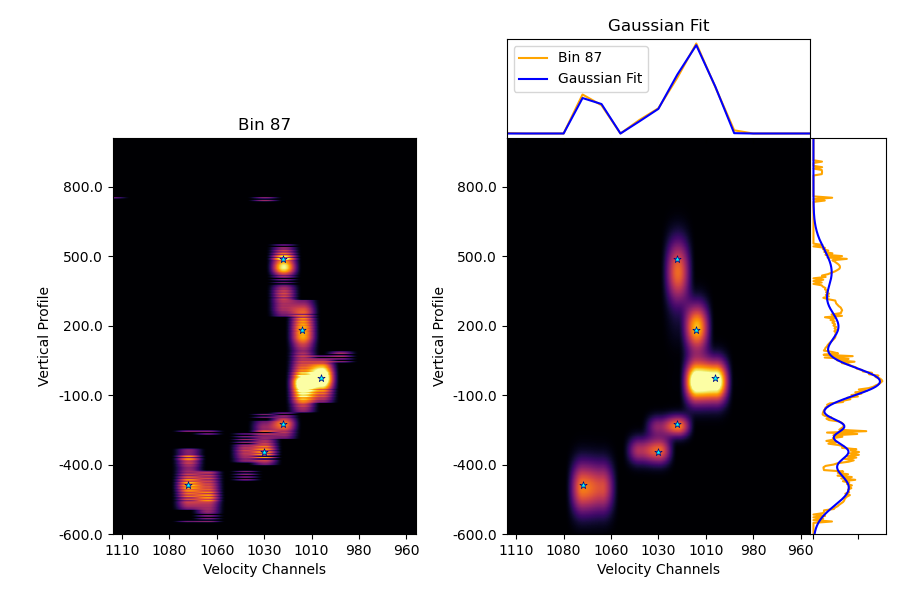}
    \caption{[Left ] White rectangles show the major ($x$) axis bins used in our peak identification and Gaussian fitting. Cyan stars show the locations of peaks identified in position velocity space (using the bin center as the $x$ coordinate). [Middle] $y$-$v$ position velocity (PV) diagram for an example bin. The PV diagram is constructed by summing the data cube along the $x$ dimension within the bin. [Right] Model PV diagram based on our fitted Gaussians. The one dimensional position and velocity profiles are also shown with the data in orange and the summed of Gaussians in the model in blue. The identified peaks appear as cyan stars on both plots.}
    \label{fig:bins}
\end{figure*}

To capture this complexity, we use a multi-Gaussian fitting method in position-velocity space.
We divide the data cube into spatial bins along the galaxy's major ($x$) axis, as illustrated in Figure~\ref{fig:bins}. Each bin is then collapsed into a two dimensional position-velocity diagram by summing along the $x$-axis within the bin. Then, we find the local maxima in the each position-velocity diagram (Figure \ref{fig:bins}) using \texttt{maximum\_filter} from \texttt{SciPy} with a sliding window of 30 square pixels corresponding to 2 beam lengths $\times$ 3 channel widths ($120$~pc $\times$ $7.5$~km~s$^{-1}$). These peaks, indicated by cyan stars in Figure \ref{fig:bins}, represent discrete structures in position–velocity space that we consider likely to correspond to individual giant molecular clouds. Their spatial positions are shown in the left panel of Figure~\ref{fig:bins}, which illustrates that individual features (clouds) often extend across several adjacent slabs.

For each $x$-axis slab, we fit two-dimensional Gaussians to all of the identified peaks at once. We model the emission using the functional form:

\begin{equation}\label{eqn:gauss}
    A_i \cdot \exp \left [- \left ( \frac{(v_i-v_{0,i})^2}{2\sigma_{v,i}^2} 
    + \frac{(y_i-y_{0,i})^2}{2\sigma_{y,i}^2} \right )\right ]
\end{equation}

\noindent where \( v \) and \( y \) are the velocity and minor axis position coordinates (i.e., the axes in Figure \ref{fig:bins}). \( A \) is the peak amplitude in units of K pc, \( v_0 \) and \( y_0 \) denote the central velocity and minor axis position of the component, \( \sigma_y \) and \( \sigma_v \) correspond to the $1\sigma$ minor axis ($y$) and velocity extent.

The right panel in Figure \ref{fig:bins} shows that these fits do a good job of capturing the complexities present in the vertical profiles. They also do a reasonable job of matching the components one would pick out by eye.


\paragraph{Interpreting the Gaussian fits}

Due to the slightly less than edge on orientation of NGC 4565, the minor axis $y$ position of any given Gaussian can either be interpreted as the vertical height above the midplane ($z$), a displacement from the major axis within the disk along the foreshortened direction parallel to the minor axis, or some combination of the two. This ambiguity means that we cannot place these features relative to a theoretical midplane based on geometry alone. We do calculate the velocity-derived radius, ($R_{\rm vel}$), for each structure. 

Similarly, \( \sigma_y \) formally corresponds to the combination of vertical and foreshortened in-plane extent of the feature being fit. For $\sigma_y$ of individual features, we expect the true vertical extent to dominate the measurement because we do not expect clouds to be $\gtrsim 1/\cos 87.5^{\rm \circ} \approx 23$ times more extended in-plane then vertically (we would see this via the accessible $x$ dimension). We discuss this more in Sections \ref{sec:incl} and Appendix \ref{sec:cprops}.


The fitted \( \sigma_v \) represents the velocity dispersion of the structure into the sky, which is oriented mostly into the plane of the galaxy. This may have some contribution from in-plane bulk motions, but we expect our Gaussian fitting approach to mostly capture the turbulent motions on the scale of individual clouds. We examine this more in Section \ref{sec:incl}. 


Considering each fitted component to reflect an individual molecular cloud, we use the fitted spatial and velocity widths, $\sigma_{v,fit}$ and $\sigma_{y,fit}$, and the amplitude, $A_{fit}$, to calculate the properties of that cloud. The linewidths are deconvolved to account for the channel width via 

\begin{equation}\label{eqn:vd}
    {\sigma_v}=\sqrt{\sigma_{v,fit}^2- \sigma_{v,chan}^2}.
\end{equation} 

\noindent Here, $\sigma_{v,chan}=0.412\Delta v$ (Equations 7 \& 8 from \citealt{10.1093/mnras/stab085}) where $\Delta_v$ is the channel width. We scale the vertical size to calculate effective radii via:

\begin{equation}\label{eqn:rdc}
    \sigma_y=\sqrt{\sigma_{y,fit}^2-(\theta_{beam}/2.355)^2},
\end{equation} 
\begin{equation}\label{eqn:reff}
    R_y=\eta \sigma_y,
\end{equation} 
where $\theta_{beam}$ is the FWHM beam size, and $\eta=1.91$ is the geometric factor to scale the measured spatial second moment to a radius for some assumed geometric profile. For convenience, we follow the classic assumption by \citet{1987ApJ...319..730S}, though we note that following \citet{2021MNRAS.502.1218R} at our resolution a Gaussian profile and a lower $\eta$ may be more appropriate.

We also integrate the Gaussians to obtain masses and densities of the objects such that

\begin{equation}
    L_{CO}^{(2-1)} = A \sigma_y \sigma_v,
\end{equation}

\begin{equation}
    \Sigma = \frac{M_{\rm lum}}{\text{Area}}=\frac{\alpha_{\rm CO}^{(2-1)}L_{\rm CO}^{(2-1)}}{\pi R_y\Delta_x},
\end{equation}

\noindent where $\alpha_{CO}^{(2-1)}$ is the same radially varying conversion factor from Section \ref{sec:conversions}, $L_{CO}^{(2-1)}$ is the line integrated luminosity of the Gaussian object in K~km~s$^{-1}$~pc$^2$ \citep{2013ARA&A..51..207B,2021MNRAS.502.1218R}, $R_y$ is the cloud size in the $y$ direction, and $\Delta_x$ is the $x$ width of bin.

Based on these properties, we also estimate the virial parameter, $\alpha_{\rm vir}$, which measures the balance between the gravitational and kinetic energy in the cloud 
\begin{equation}
    \alpha_{vir}=\frac{5\sigma_v^2 R}{GM}\propto\frac{\sigma_v^2}{\Sigma R}
\end{equation}
where $G$ is the gravitational constant and $\alpha_{vir}=1$ for gas in virial equilibrium assuming only self-gravity and a fixed density sphere geometry. 

\paragraph{\texttt{CPROPS}-based molecular cloud properties}

To check our results, we also apply the \texttt{CPROPS} algorithm to our data and analyze the output in Appendix \ref{sec:cprops}. 
We find that the two methods largely agree with one another, although \texttt{CPROPS} tends to select slightly larger structures. This supports the rigor of the comparison below (\S \ref{sec:env} and \ref{sec:incl}) between our measurements and \texttt{CPROPS}-based analysis of M31 $^{12}$CO emission \citep{2023MNRAS.522.6137P} and the 60 pc resolution PHANGS $^{12}$CO(2-1), both derived using \texttt{CPROPS}. \texttt{CPROPS} also measures the orientation of the structures that it finds. We find that the clouds in NGC 4565 are mostly -- but not entirely -- aligned with the major axis of the galaxy (Section \ref{sec:incl}).

\subsection{Vertical Thickness of the Molecular Gas Layer}\label{sec:hmol}

\begin{figure*}
    \centering
    \includegraphics[width=.8\textwidth]{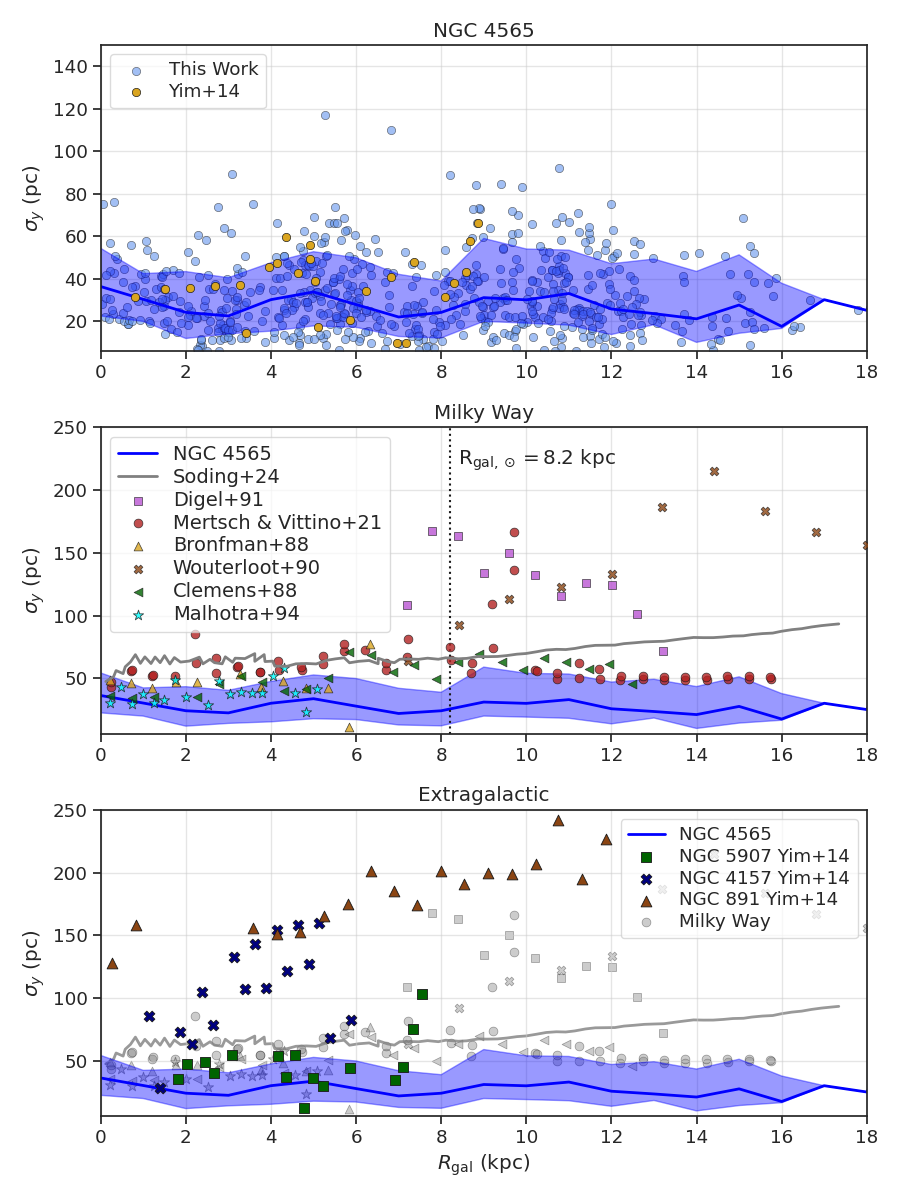}
    \caption{[Top] Our fitted $\sigma_y$ of structures in NGC 4565 with median $\sigma_y$ and 16th-84th percentiles (shaded) are shown in dark blue. The CO scale height from \citet{2014AJ....148..127Y} is shown for comparison. [Middle] Scale heights for CO in NGC 4565 compared to the Milky Way. [Bottom] Scale heights of CO for NGC 5907, NGC 4157, and NGC 891 compared to NGC 4565.}
    \label{fig:hmol}
\end{figure*}

In Figure \ref{fig:hmol} we plot the minor axis extent of molecular clouds ($\sigma_y$) as a function of $R_{\rm vel}$. We compare our results to other measurements of the vertical thickness of the molecular gas layer in NGC 4565, the Milky Way, and several other nearby galaxies. 

The $\sigma_y$ that we derive appear almost constant as a function of galactocentric radius. We emphasize that this is the \textit{vertical thickness of individual components}. Our method does not estimate the vertical distribution of those components about the midplane. If we were to observe NGC 4565 face-on at high resolution, then our measurement would correspond to the correct line-of-sight depth to translate cloud-by-cloud surface densities into volume densities. But our measurement does not indicate, e.g., whether the individual clouds are scattered vertically throughout the extended atomic gas layer.

Fig. \ref{fig:hmol} compares our measurements to previous work. \citet{2014AJ....148..127Y} used a modified version of the double Gaussian method following \citet{1996AJ....112..457O} to determine the scale height of the molecular and atomic gas in NGC 4565, NGC 5907, NGC 4157. In NGC 891, \citet{2011AJ....141...48Y} used a standard Gaussian fitting method because that galaxy is almost perfectly edge-on. The  CARMA data used in these studies data has less sensitivity and coarser resolution than our ALMA data, and their methods emphasize larger scales than our component-by-component fitting. Even with these caveats, \citet{2014AJ....148..127Y} find a similar mean scale height measurement for NGC 4565, although our component-by-component fitting results in a larger scatter. In Figure \ref{fig:hmol}, their measurements overlap ours and they also find a nearly flat gradient of $\sigma_{y}$ vs. $R_{\rm gal}$, only 1.1 pc/kpc. 

We also compare $\sigma_y$ vs. $R_{\rm gal}$ in NGC 4565 to measurements from the Milky Way (Figure \ref{fig:hmol}; middle). Galactic molecular gas scale heights have been derived using all of the previously mentioned methods. Within the solar circle \citep[8.2 kpc;][]{2023MNRAS.519..948L}, different methods yield consistent results. Most measurements suggest a thin molecular gas disk with $\sigma_y \approx 39{-}64$~pc, only slightly larger than what we measure for NGC 4565. In the outer Galaxy, some measurements remain thin while some indicate significant flaring of the molecular disk with increasing $R_{\rm gal}$. These differences likely reflect the different methods and different studies targeting different parts of the disk \citep{2015ARA&A..53..583H}. Given this internal scatter in Galactic measurements, we do not draw any strong conclusions comparing NGC 4565 to the outer Galaxy.


Looking at other highly inclined galaxies (Figure \ref{fig:hmol}; Bottom), there appear to be two distinct trends in $\sigma_y$. For NGC 4565, NGC 5907, and the Milky Way, the molecular scale heights are small and do not flare much with $R_{\rm gal}$. Meanwhile, the scale heights for NGC 891 and NGC 4157 are larger and increase with $R_{\rm gal}$. This is similar to the vertical flaring with radius that atomic disks exhibit due to a weaker potential well and lower dynamical equilibrium pressure in the outer parts of galaxies \citep{2022ApJ...936..137O}.

The star formation rate surface density, which relates to the intensity of stellar feedback, has been suggested as a factor that contributes to these differences in the vertical structure of molecular gas. NGC 4565, NGC 5907, and the Milky Way all have relatively low SFR surface densities compared to NGC 891 and NGC 4157. This could result in thinner molecular disks due to reduced turbulent support. \citet{2014AJ....148..127Y} found a weak correlation between the galaxy-averaged CO disk thickness ($h_{\rm mol}$) and the SFR among these galaxies, although they found $h_{mol}$ and $\Sigma_{SFR}$ to be anti-correlated within a galaxy. Consistent with this picture, there is little or no evidence of galaxy-scale winds in NGC 4565 and NGC 5907 \citep{2019A&A...632A..12S,2026A&A...706A.374S}, while NGC 891 and NGC 4157 both have been found to host galactic wind or fountain activity \citep{2021MNRAS.502..969Y,2024A&A...690A.348C,2026A&A...706A.374S}. A more detailed treatment would balance the stellar feedback against the local disk potential, but it seems reasonable that galaxies with stronger feedback show larger scale heights and apparently this may also correspond to radially increasing molecular scale heights.


\begin{deluxetable*}{l|ccc|ccc|ccc|cc}[!htb]
\tablecaption{Average GMC Properties and Radial Trends\label{tab:combined_gmc}}
\tablehead{
\colhead{Galaxy} & 
\multicolumn{3}{c}{$\sigma_v$ (km s$^{-1}$)} & 
\multicolumn{3}{c}{$\Sigma$ ($M_\odot$ pc$^{-2}$)} & 
\multicolumn{3}{c}{$\alpha_{\rm vir}$} & 
\colhead{${\sigma_y}$ (pc)} \\
\colhead{} & 
\colhead{Avg} & \colhead{Spearman $r$} & \colhead{Slope\tablenotemark{a}} &
\colhead{Avg} & \colhead{Spearman $r$} & \colhead{Slope} &
\colhead{Avg} & \colhead{Spearman $r$} & \colhead{Slope} &
\colhead{}
}
\startdata
NGC 4565   & 5.3$\pm$0.1 & -0.424 & -0.027 & 91$\pm$4.5 & -0.122 & -0.029 & 3.91$\pm$0.40 & -0.630 & -0.054 & 33.6$\pm$0.7 \\
Milky Way  & 5.4$\pm$0.1 & -0.574 & -0.009 & 132$\pm$5 & -0.524 & -0.050 & 4.9$\pm$1.9 & -0.175 & 0.025 & 66.8$\pm$3.2\tablenotemark{b} \\
PHANGS     & 4.2$\pm$0.1 & -0.390 & -0.043 & 177$\pm$3 & -0.180 & -0.007 & 2.2 & -0.300 & -0.070 & 31.2$\pm$0.2 \\
M31        & 4.9 & ... & ... & 54$\pm$31 & ... & ... & 5.4$\pm$2.2 & ... & ... & 34.6$\pm$1.0 \\
NGC 5907   & ... & ... & ... & ... & ... & ... & ... & ... & ... & 47.2$\pm$4.9 \tablenotemark{c}\\
NGC 4157   & ... & ... & ... & ... & ... & ... & ... & ... & ... & 106$\pm$9 \tablenotemark{c}\\
NGC 891    & ... & ... & ... & ... & ... & ... & ... & ... & ... & 184$\pm$7 \tablenotemark{c}\\
\enddata
\tablenotetext{a}{dex kpc$^{-1}$}
\tablenotetext{b}{Milky Way $\sigma_y$ from \citet{2017ApJ...834...57M}; alternate value from Eq.~\ref{eqn:reff} is $24.8\pm0.3$ pc.}
\tablenotetext{c}{\citet{2011AJ....141...48Y}}
\end{deluxetable*}

\subsection{Radial Trends in Cloud Properties}\label{sec:gmccomp}
Other $^{12}$CO observations reach the $60$~pc scales that we consider. These tend to show that physical properties of molecular clouds --- including cloud sizes ($R$), velocity dispersions ($\sigma_v$), surface densities ($\Sigma_{\rm mol}$), and virial parameters ($\alpha_{vir}$) --- show lognormal distributions \citep{2018ApJ...860..172S,2020ApJ...901L...8S,2021MNRAS.502.1218R} and correlate with environmental factors like stellar surface density and large-scale average gas surface density \citep{2022AJ....164...43S}. However, these studies tend to target systems viewed at low inclination and focus on the inner disks of galaxies. Our observations of NGC 4565 offer a high inclination view that extends to large galactocentric radius and captures a wide range of surface densities (Section \ref{sec:radial_analysis}).

We compare our results to the subset of the PHANGS--ALMA sample that achieves 60 pc resolution \citep{2021MNRAS.502.1218R,2022AJ....164...43S}, which are more face-on compared to NGC 4565. To augment the high inclination sample, we also compare to GMC catalogs for the Milky Way \citep{2017ApJ...834...57M}, which we view with a nearly edge-on perspective and M31 \citep{2016AJ....151...34C,2016ApJ...831...16L,2019ApJ...883....2S}, which also has high $\approx 78^\circ$ inclination. While M31 and the PHANGS sample match the resolution of our NGC 4565 data, the Milky Way observations are more heterogeneous due to the nature of observing from within the Galaxy. In order to produce a more homogeneous sample of Milky Way clouds, we add a fixed FWHM $60$~pc beam size in quadrature to the measured cloud sizes and then recompute surface densities by dividing the cloud masses by these beam-diluted areas. We then select only the Milky Way clouds that we would expect to observe in NGC 4565 after applying a threshold to this beam-diluted surface density that corresponds to our sensitivity limit in NGC 4565. After this selection, we analyze the original (i.e., not artificially beam diluted) values from the Milky Way catalog. 

\begin{figure*}
    \centering
    \vspace{-.2in}
    \includegraphics[trim={.48cm 1.6cm .35cm 2.4cm},clip,width=0.9\textwidth]{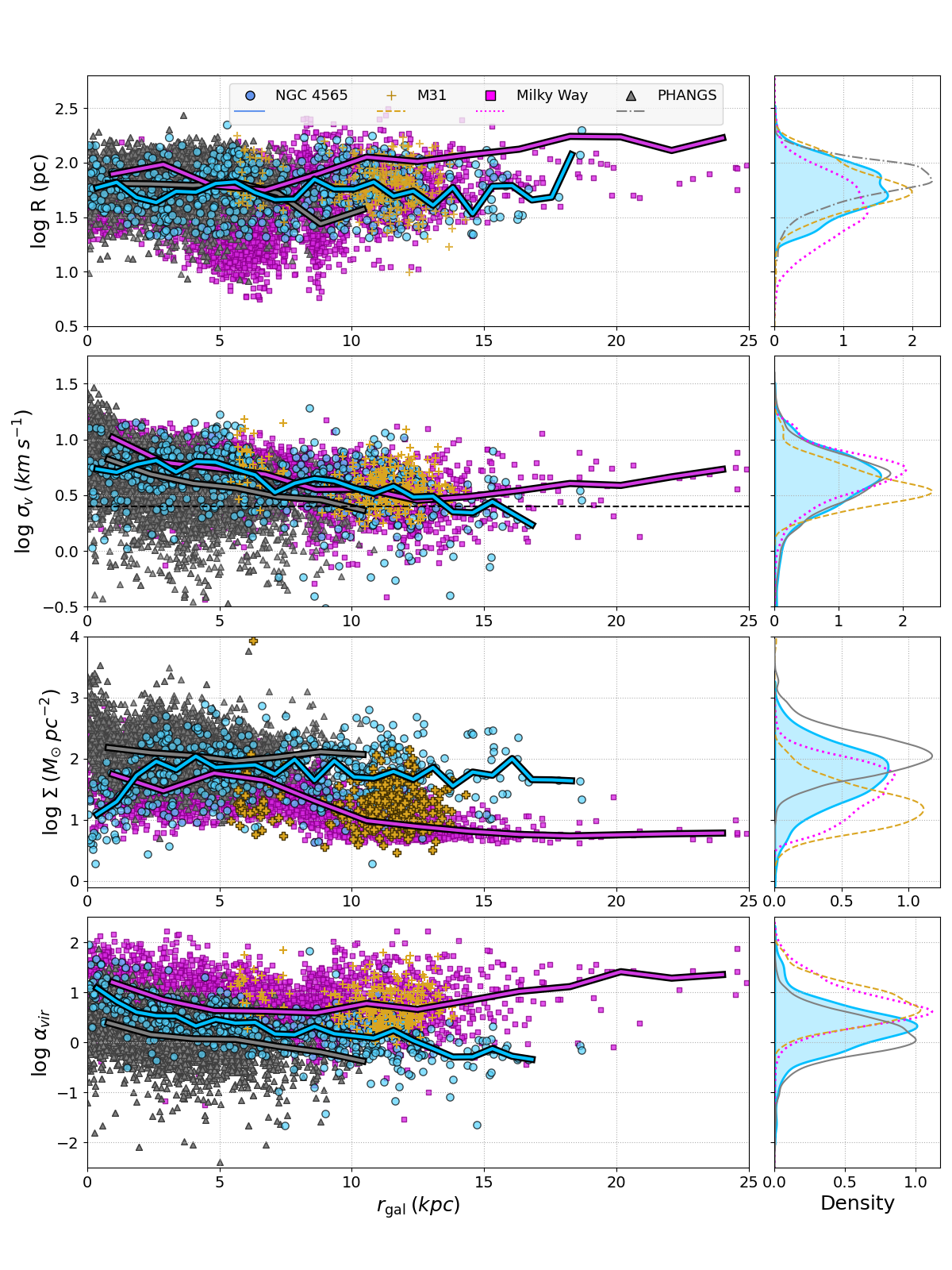}
    \caption{GMC sizes, velocity dispersions, surface densities, and virial parameters as a function of galactocentric radius. The binned averages are represented as solid colored lines for each galaxy. The distributions of the properties are represented as kernel density estimates (KDEs) to the right. In the $\sigma_v$ plot, the 2.5 km/s channel width is shown with a dashed black line.}
    \label{fig:prop_kde}
    \vspace{-.2in}
\end{figure*}

Figure \ref{fig:prop_kde} shows cloud sizes, velocity dispersions, surface densities, and virial parameters as a function of galactocentric radius, as well as their overall distributions. Following Equation \ref{eqn:reff}, the sizes for NGC 4565 reflect only the vertical extent of the cloud. We plot results for each fit component in each slice, so clouds that extend across multiple position-velocity slices (see above) may contribute multiple data points. We report the distributions of properties and their radial gradients in Table \ref{tab:combined_gmc}.

For cloud sizes, all samples except the Milky Way show similar distributions and no significant trend in galactocentric radius (Table \ref{tab:combined_gmc}). Most cloud-finding algorithms are biased towards finding beam-sized structures \citep{pineda_2009,2013ApJ...779...46H,2016ApJ...831...16L}, and this drives the similarity in the matched resolution measurements for NGC 4565, PHANGS--ALMA, and M31. The Milky Way exhibits a wider distribution that includes smaller clouds near the solar circle and increasing radius with distance. This just reflects the heterogeneous physical resolution of Galactic data, which depends on the location in the galaxy. Because of the concentration of M31 clouds into the 10-kpc ring \citep{2006A&A...453..459N}, we do not measure radial gradients for that galaxy.

GMC linewidths also show mostly similar distributions across galaxies. 
For NGC~4565, PHANGS-ALMA, and the Milky Way, we find a decrease in linewidth with increasing radius, which ranges from 0.01-0.04 dex kpc$^{-1}$ (Table \ref{tab:combined_gmc}). We see in the next section that the lower line width in M31 appears to reflect the location of those molecular clouds far from the center of that galaxy. 

Both the surface densities and virial parameters also decrease with increasing $R_{\rm gal}$. For PHANGS--ALMA the radial gradient in surface density appears less significant, but we will see in the next section that this partially reflects blending many different galaxies together. The declining virial parameter mirrors the decline in the line width. The higher virial parameters and lower surface densities of the Milky Way and M31 clouds may suggest either that many of these clouds remain unresolved (so that their surface density is an underestimate) or that they are confined by ISM pressure rather than bound by self-gravity \citep[][]{2019ApJ...883....2S}. Again the Milky Way measurements also reflect the different and heterogeneous resolution of those data.

\begin{figure}
    \centering
    \includegraphics[width=0.49\textwidth]{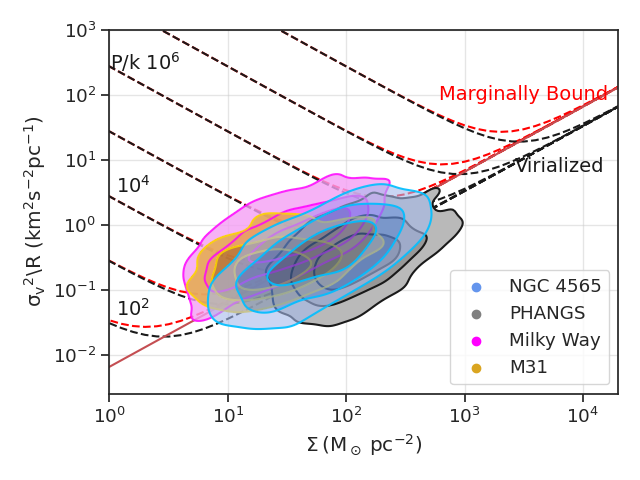}
    \caption{Heyer-Keto relation for molecular clouds in NGC 4565 plotting the velocity dispersion normalized by the cloud size, $\mathrm{\sigma_{v}^{2}/R}$, as a function of molecular gas surface density, $\Sigma$. The black lines represent virial equilibrium and the red lines represent marginally boundedness. The dashed black and red lines represent those same conditions accounting for external pressure ranging from $10^{2}-10^{8}$ K $\mathrm{cm^{-3}}$ \citep[][]{2011MNRAS.416..710F}}
    \label{fig:heyer}
\end{figure}


The similarity of the distributions and radial trends suggests that at 60~pc resolution clouds viewed face-on in PHANGS--ALMA resemble clouds viewed edge-on in NGC~4565. To test this directly, in Figure \ref{fig:heyer} we plot all samples on the ``Heyer-Keto'' diagram, the relationship between turbulent kinetic energy ($\sigma_v^2/R$) and gravitational potential energy ($\Sigma$) following \citet{article2,2024ARA&A..62..369S}. The location of clouds in this space indicates the extent to which they are bound by gas self-gravity. The plot also captures variations in surface density and turbulent pressure within clouds.

In Figure \ref{fig:heyer}, all of the datasets tend to fall around the marginally bound and virialized lines. As indicated in Figure \ref{fig:prop_kde} the Milky Way and M31 measurements show higher $\alpha_{\rm vir}$ on average, and so appear higher in this plot. This may indicate that these clouds are confined by modest amounts ($\sim10^3-10^6\text{ K cm}^{-3}$) of external pressure or harbor significant mass in other components (e.g., atomic gas). Alternatively, it may reflect some of the systematic effects mentioned above. For M31 an important role for external pressure shaping cloud properties has been noted by \citet{2019ApJ...883....2S}.

Returning to our original goal, the properties of clouds viewed edge-on in NGC 4565 appear similar, even indistinguishable, from clouds observed at similar resolution in more face-on galaxies in PHANGS--ALMA. This is true despite the fact that we only use the vertical size, $\sigma_y$, in these calculations. We return to the question of whether clouds appear truly isotropic in Section \ref{sec:incl} but to first order the edge-on clouds in NGC 4565 appear ``normal.''

\subsection{Cloud Properties and Environment}\label{sec:env}



The properties of molecular clouds are thought to reflect the larger-scale conditions of galaxy disks. Variations in the ambient pressure, gravitational potential, and mean density of the interstellar medium can shape the characteristic density and turbulent state of molecular gas, leading to systematic differences in cloud populations across galactic environments \citep{2022AJ....164...43S,2024ARA&A..62..369S}.

Figure \ref{fig:prop_kde} shows radial gradients in cloud properties that reflect a link between GMC properties and external conditions. In \citet{2022AJ....164...43S}, cloud properties correlate with many properties of the larger scale galaxy disk including kpc-scale molecular gas surface density, $\Sigma_{\rm mol,kpc}$, stellar surface density, $\Sigma_{\rm*,kpc}$, and global properties like the star formation rate or total molecular gas mass. Among these, many of the strongest correlations link molecular cloud properties to the large ($\approx 1$~kpc) scale molecular gas surface density of the disk, $\Sigma_{\rm mol,kpc}$. In turn $\Sigma_{\rm mol,kpc}$ relates to the dynamical equilibrium pressure \citep{2020ApJ...892..148S,2022ApJ...936..137O} and mean density of the ISM.



\begin{figure*}
    \centering
    \includegraphics[trim={0cm 1.5cm 0cm 1.2cm},clip,width=0.8\textwidth]{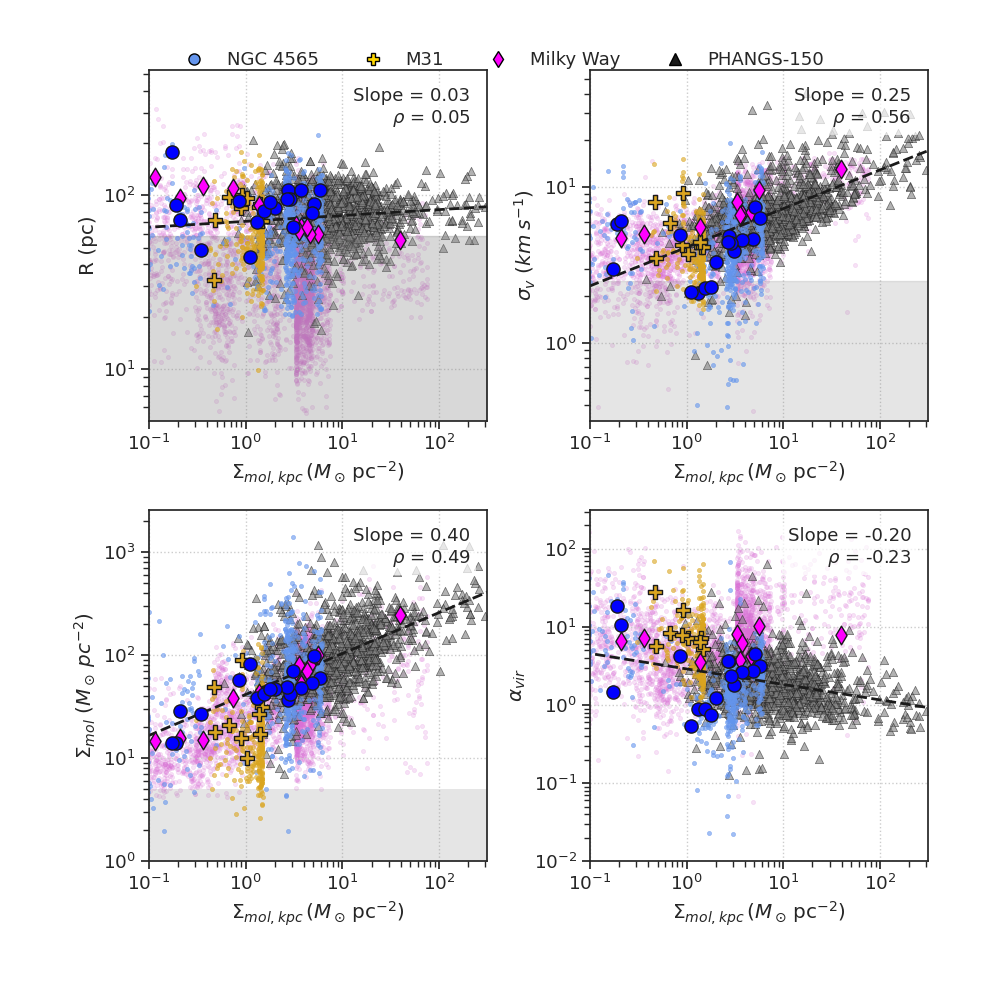}
    \caption{GMC R (perpendicular), $\sigma_v$, $\Sigma$, and $\alpha_{vir}$ vs. the local environmental (i.e., large scale) molecular gas surface at the $R_{gal}$ of the GMC. The square markers and gray triangles represent the mass-weighted average in each kpc-wide radial bin. The black dashed line represents the best fit line of the binned properties with the fitting range limited to $\Sigma_{\rm mol,kpc}>\rm 10^{0}\:M_\odot\:pc^{-2}$ to avoid sensitivity biases. The gray shaded regions reflect the NGC 4565 FWHM spatial and spectral channel width, as well as the surface density sensitivity limit. While these are not hard limits, points that fall in these shaded regions should be considered marginal. For all subplots, the GMC properties remain uncorrected by inclination while the large-scale $\Sigma_{\rm mol,kpc}$ is the face-on equivalent either by reconstructing from a stripe integral-type modeling (for NGC 4565 or the Milky Way) or by using a correction factor of $\cos(i)$ (the other galaxies).}
    \label{fig:props_env}
\end{figure*}

In Figure \ref{fig:props_env}, we plot GMC properties as a function of the large scale molecular gas surface density, $\Sigma_{\rm mol,kpc}$. For NGC 4565 we draw $\Sigma_{\rm mol,kpc}$ from the radial profiles in Figure \ref{fig:radprof}. Here we use the full PHANGS catalog from \citet{2022AJ....164...43S} at 150 pc resolution with 0.5 kpc rings to obtain a wider sample of physical properties. For NGC 4565, M31, and the Milky Way, we aggregate cloud properties within a region following a procedure similar to \citet{2022AJ....164...43S}. We separated each galaxy into 1~kpc-wide radial bins. Then within each bin, we compute mass-weighted averages of the GMC properties. For the PHANGS sample, we used the mass-weighted averages of \texttt{CPROPS} cloud properties calculated in radial bins by \citet{2022AJ....164...43S}.

In good agreement with \citet{2022AJ....164...43S}, we find $\sigma_v$ and $\Sigma_{\rm mol}$ to have strong correlations with $\Sigma_{\rm mol,kpc}$, while $R$ and $\alpha_{\rm vir}$ have weaker correlations. 
NGC 4565 generally falls along the same trends as less-inclined PHANGS galaxies, as does M31 for the most part \citep[for M31 see also][]{2019ApJ...883....2S}. In detail, the correlation with line width may be somewhat steeper for NGC 4565 than the PHANGS-ALMA 150~pc data, with very low line widths found in the outer disk of NGC 4565. The most notable difference between PHANGS-ALMA and NGC 4565 result is that the virial parameter in NGC 4565 appears to decrease with increasing radius and so to increase with increasing surface density over the range $\Sigma_{\rm mol, kpc} \approx 1{-}10$~M$_\odot$~pc$^{-2}$. The overall anti-correlation between $\alpha_{\rm vir}$ and $\Sigma_{\rm mol, kpc}$ obscures this, but reflects the handful of low surface density bins with high line widths from the inner galaxy (see also Figure \ref{fig:radcomp}). The Milky Way broadly shows similar trends, though with line widths offset to higher values at fixed $\Sigma_{\rm mol, kpc}$ and clear effects of our imposed surface brightness cut in the $R$ and $\Sigma_{\rm mol}$ trends. 

Figure \ref{fig:props_env} demonstrates that the cloud properties in NGC 4565 show a similar coupling to large scale environment as a large sample of other nearby galaxies, though with a few important caveats. It also shows that high resolution observations of highly inclined galaxies may be an efficient way to extend these studies to a wider range of galactic environments.




\subsection{Cloud Properties as a Function of Inclination\label{sec:incl}}

It is common to assume isotropic geometry when describing molecular cloud properties. However, molecular clouds may display asymmetry between their in-plane and vertical extent, velocity dispersion, and projected surface density. Following Figure \ref{fig:radcomp} the molecular gas layer itself is clearly thin.  \citet{2021MNRAS.502.1218R} and \citet{2022AJ....164...43S} assume that the large objects identified as clouds in PHANGS--ALMA are oblate and flattened to have $\leq 100$~pc vertical extent. Analyzing PHANGS--ALMA at 60-150 pc resolutions, \citet{2022AJ....164...43S} and \citet{HughesInPrep} found systematic trends with galaxy inclination in both \texttt{CPROPS}-based and pixel-by-pixel cloud-scale gas property measurements. To account for these, \citet{2022AJ....164...43S} proposed corrections to account for the effect of inclination  on cloud properties. These assume an oblate geometry with the shortened dimension along the vertical axis and modify the observed quantities as follows:


\begin{equation}\label{eqn:sigv_incl}
    \sigma_v = \sigma_{v,obs}(\cos(i))^{0.5}
\end{equation}

\begin{equation}\label{eqn:Sig_incl}
    \Sigma = \Sigma_{obs}\cos(i).
\end{equation}

\noindent Equation \ref{eqn:sigv_incl} is empirically-based, while Equation \ref{eqn:Sig_incl} is the geometric correction for an inclined thin disk. 

More generally, a spheroidal cloud may have some vertical extent and some in-plane extent. Its velocity dispersion may similarly have a vertical and an in-plane motion. Then the observed velocity dispersion is 

\begin{equation}\label{eqn:sigobs_geom}
    \sigma_{obs} = \sigma_\| \sqrt{\eta_\sigma^2\cos^2 i + \sin^2i},
\end{equation}

\noindent where $\eta_\sigma$ is the ratio of the in-plane and vertical velocity dispersions such that $\eta_\sigma=\sigma_{\perp}/\sigma_{\|}$. For the observed radius, assuming spheroidal geometry gives a similar relationship

\begin{equation}\label{eqn:robs_sph}
    R_{obs} = R_\| \sqrt{\cos^2 i + \eta_R^2\sin^2i}.
\end{equation}

\noindent where $\eta_R = R_{\perp}/R_{\|}$. This means that for a completely face-on galaxy, $\sigma_{obs,i=0}=\sigma_{\perp}$ and $R_{obs,i=0}=R_{\|}$, while the opposite is true for an exactly edge-on galaxy where $\sigma_{obs,i=90}=\sigma_{\|}$ and $R_{obs,i=90}=R_{\perp}$. We can also use the geometric corrections for $R_{obs}$ and $\sigma_{obs}$ to predict geometric effects on $\Sigma$ and $\alpha_{vir}$ given some $\eta_R$ and $\eta_v$.

Contrasting cloud properties found in similar environments in highly inclined galaxies and face-on galaxies offers empirical constraints on $\eta_{\sigma}$ and $\eta_{R}$. Highly inclined galaxies also offer the chance to estimate $\eta_R$ directly by contrasting the in-plane and vertical extent of measured structures. We adopt both approaches.

\paragraph{Direct measurement from cloud properties}

Using the outputs from \texttt{CPROPS} (Appendix \ref{sec:cprops}), $\eta_R$ can be estimated from the ratio of cloud sizes along the major and minor axes, such that $\eta_{ab} = R_{\rm b} / R_{\rm a}$, where $R_{\rm a}$ and $R_{\rm b}$ are the sizes along the $a$ (major) and $b$ (minor) axes of the cloud. We show the axis ratios ($1/\eta_{ab}$) and orientation of clouds identified by \texttt{CPROPS} in Figure \ref{fig:res_incl}. Using this approach, we find a median $\eta_{ab}$ for NGC 4565 to be 0.67 corresponding to axis ratio $\approx 1.49$. We treat this $\eta_{ab}$ as a proxy for $\eta_R$ in Figure \ref{fig:res_incl} due to the the identified clouds being preferentially aligned with the major axis of the galaxy. 

\begin{figure}
    \centering
    \includegraphics[width=0.48\textwidth]{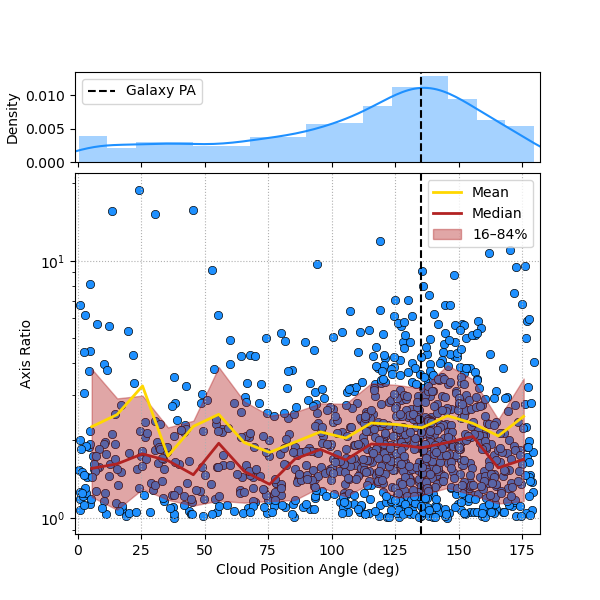}
    \caption{[Top] Cloud position angles estimated using \texttt{CPROPS} measured north through east. The orientation of the major axis of the galaxy (PA = 135$^\circ$) is shown with a dashed line. 67.7\% of cloud position angles are within 30$^\circ$ of the galaxy position angle. [Bottom] Ratio of the major to minor axis size of clouds as a function of their position angles. The median and mean axis ratio for NGC 4565 across all position angles are 1.49 and 2.2 respectively.}
    \label{fig:cprops_posang}
\end{figure}

\begin{figure*}
    \centering
    \includegraphics[trim={0cm 0cm 0cm 1.2cm},clip,width=0.8\linewidth]{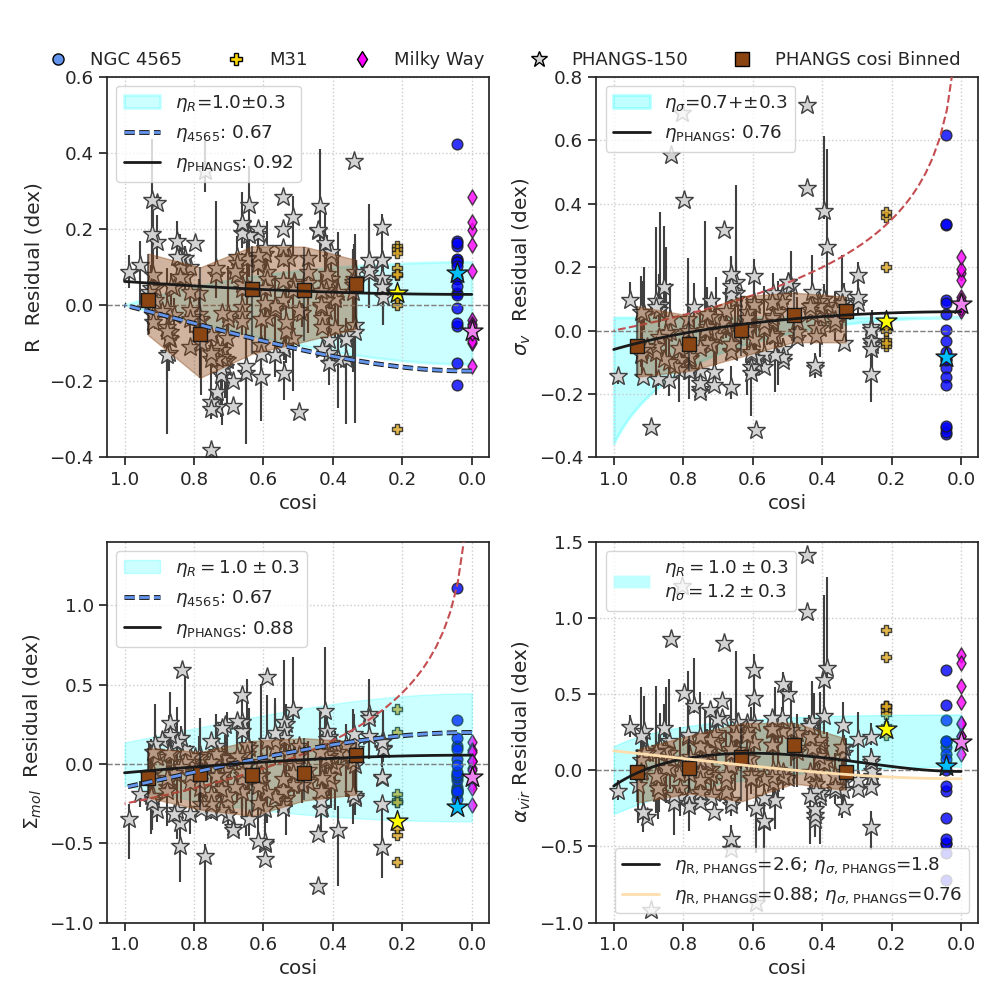}
    \caption{Residuals about the relationships between weighted mean cloud properties and $\Sigma_{\rm mol,kpc}$ from Figure \ref{fig:props_env} as a function of galaxy inclination, parameterized by $\cos{i}$. For the PHANGS galaxies (gray stars), the brown shaded region and square points show the median and 16-84\% range averaging all rings in all galaxies together in $0.15$-wide bins in $\cos i$. The dashed red lines represent the expected trends with inclination from Equations \ref{eqn:sigv_incl} and \ref{eqn:Sig_incl}. The dashed blue lines show the expectations from the \texttt{CPROPS}-based $\eta$. The shaded regions show geometric inclination effects from Equations \ref{eqn:sigobs_geom} and \ref{eqn:robs_sph} for a range of $\eta$.}
    \label{fig:res_incl}
\end{figure*}

This preferential alignment was found with the \texttt{CPROPS} estimates the cloud position angle, which we show in the top panel in Figure \ref{fig:cprops_posang}. We find 60.7\% of cloud position angles are within 30$^\circ$ of the galaxy position angle, though there is some population of clouds oriented at every position angle. Based on the bottom panel, the axis ratio also does not appear to be correlated with cloud position angle. 

An extension of this would could contrast the axis ratios of clouds in highly inclined galaxies to those in more face-on galaxies. This would test whether inclination is primarily an alignment of filaments,
as results in the Milky Way suggest \citep[e.g.,][]{2018ApJ...864..153Z,2019ApJ...887..186Z}, or oblate spheroids.

\paragraph{Statistical measurement from comparing galaxies} We also test how deviations from observed cloud-environment scaling relation correlate with inclination. For cloud catalogs from other galaxies, we recalculated the radius $R$ to include only the component perpendicular to the major axis of the galaxy\footnote{Typically \texttt{CPROPS} catalogs express cloud radius as the geometric mean of the cloud major and minor axes.}.

To control for the variation of cloud properties with environment, we calculated the residuals of each physical property relative to the best-fit relations presented in Figure \ref{fig:props_env}. Then in Figure \ref{fig:res_incl} we examine trends in these residuals as a function of 
$\cos(i)$.

Figure \ref{fig:res_incl} shows residuals about cloud property$-\Sigma_{\rm mol,kpc}$ relations from Figure \ref{fig:props_env} as a function of galaxy inclination. We take the mass-weighted average of the binned residuals for each galaxy to see how the global average properties are systemically affected by galaxy inclination. The residuals for $\sigma_v$ appear to correlate more strongly with inclination than the other properties with a Spearman rank correlation coefficient of -0.38 relating $\sigma_v$ to $\cos i$, compared to correlation coefficients of -0.06, -0.11, and -0.12 for $R$, $\Sigma$, and $\alpha_{vir}$ respectively.

The empirical inclination corrections (Equations \ref{eqn:sigv_incl} and \ref{eqn:Sig_incl}) from \citet{2022AJ....164...43S} fit the low to moderately inclined galaxies, but predict a much larger effect in nearly edge-on galaxies than we observe in our data. If we instead apply our geometric framework (Equations \ref{eqn:sigobs_geom} and \ref{eqn:robs_sph}), we find best-fit $\eta_R$ and $\eta_\sigma$ values for the PHANGS galaxies to be 0.92 for $R$, 0.76 for  $\sigma_v$, and 0.88 for $\Sigma_{\rm mol}$. Similarly fitting the residuals of $\alpha_{vir}$ as a combination of dependence on $\eta_R$ and $\eta_\sigma$ gives best-fit values of $\eta_R=2.6$ and $\eta_\sigma=1.8$ for the PHANGS galaxies. However, because $\alpha_{vir}$ depends on both $\eta$ values, this does not constrain the $\eta_R$ and $\eta_\sigma$ individually and instead constrains their relation to one another. Using the independently best fit values of $\eta_R=0.88$ and $\eta_\sigma=0.76$ from the $\Sigma_{\rm mol}$ and $\sigma_v$ measurements gives a line showing a similar behavior to the best fit $\eta_R=2.6$ and $\eta_\sigma=1.8$ line at the inclination extremes, but with a different curvature in the middle. 

The spectral dimension is more affected by inclination in this test. The $\sigma_v$ residuals from PHANGS vary with inclination consistent with $\eta_\sigma=0.7\pm0.3$. This could be indicative of a small degree of broadening due to in-plane motions or with internal velocity anisotropy within the clouds themselves. Unlike $\sigma_R$, $\sigma_v$ is not easily accessible inside an edge-on galaxy since it only gives one (in--plane) dimension of velocity dispersion.

As discussed above, $R$ is most strongly impacted by the resolution of the data rather than location in the galaxy or local ISM conditions-- especially since our 55 pc resolution is the same scale as the in plane height of the clouds. It is possible that any inclination dependence in $R$ is more likely to appear indirectly through its contribution to $\Sigma_{\mathrm{mol}}$ which is less affected by the ability to vertically resolve the clouds. While the projected $R$ on the sky affects all galaxies equally, the in-plane $R$ contributes to $\Sigma_{\mathrm{mol}}$, making it possible to detect $\eta_R$ statistically even without fully resolving clouds vertically-- provided the clouds have a flattened, disk morphology rather than elongated filamentary shapes \citep{2018ApJ...864..153Z,2019ApJ...887..186Z,2022ApJ...936..160Z}.

This would suggest that the flattened appearance of the cloud population, as implied by the \texttt{CPROPS} axis ratio ($\eta_R \approx 0.67$), reflects a modest degree of vertical compression that becomes more apparent in the inclination trends of $\Sigma_{\mathrm{mol}}$ rather than in $R$ itself.

In both surface density and velocity dispersion, our NGC 4565 measurements appear to lie below the extrapolation from the PHANGS galaxies. Their position is consistent with Figure \ref{fig:props_env}, where NGC 4565 appears below the best-fit environmental trend. And the cloud properties in NGC 4565 do indicate asymmetry. But the main contribution of NGC 4565 to this statistical measurement is to confirm that edge-on cloud properties are not dramatically different than face-on ones.

While the ranges of $\eta$s do indicate a potential systemic inclination effect on the molecular cloud properties, they do not rule out the isotropic case of $\eta_R$ and $\eta_\sigma=1$, highlighting the limited power of purely statistical methods to gauge cloud anisotropy. Resolved structural analysis presents a clearer path to understanding $R$ than purely statistical measurements since it directly probes cloud geometry and vertical compression. Statistical methods remain valuable for $\sigma_v$, which has no easily accessible structural measurement.

\subsection{East Ring Pileup and the Jewel\label{sec:jewel}}

\begin{figure*}
    \centering
    \includegraphics[width=0.9\textwidth]{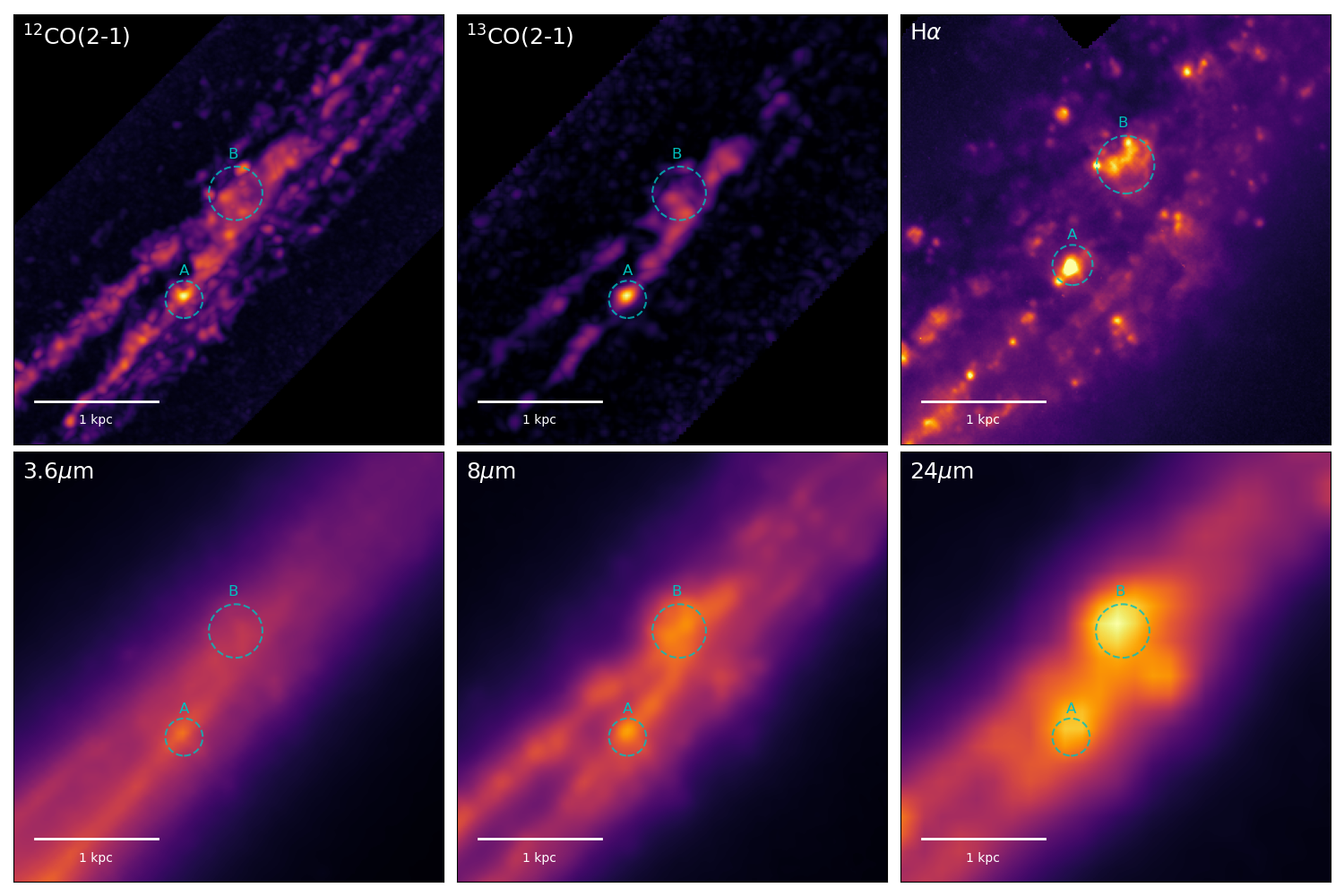}
    \caption{Multi-wavelength view of the East Ring Pileup which contains the Jewel (A) as the brightest $^{12}$CO source, as well as another region (B) which represent the two brightest H$\alpha$ and 24$\mu$m peaks. This includes our high-resolution CO observations, archival Spitzer infrared imaging used previously for the radial profiles, and newly obtained MUSE H$\alpha$ from the GECKOS survey (J. van de Sande et al. in preparation; \citealp{2024IAUS..377...27V}).}
    \label{fig:jewel}
\end{figure*}

One of the most prominent features revealed by our  observations is the bright emission in the eastern corner of NGC 4565's ring. This region has the brightest $^{12}$CO(2-1) and $^{13}$CO(2-1) in the galaxy and corresponds to a high density of molecular gas, which we refer to as the East Ring Pileup. The East Ring Pileup is shown in a multi-wavelength view in Figure \ref{fig:jewel} which demonstrates that this region contains the three brightest star forming regions in the galaxy (regions A and B). While extranuclear sites of massive star formation are often associated with mergers \citep{1996ApJ...471..115B} like the Antennae Galaxies \citep{1999AJ....118.1551W} or NGC 3256 \citep{1999AJ....118..752Z}, the East Ring Pileup in NGC 4565 occurs within an a disk that is mostly undisturbed by its satellite galaxies. This is more similar to the ``headlight cloud" in NGC 628 as an example of an extreme giant molecular complex embedded within a galactic disk \citep{2020A&A...634A.121H}.

This Pileup contains a particularly bright and compact feature that we refer to as the Jewel (Region A in Figure \ref{fig:jewel} and Table \ref{tab:jewel}). Although identified in our new CO observations, the Jewel also appears prominent across multiple wavelengths in Figure \ref{fig:jewel}. It is the brightest source in NGC~4565 at $\lambda = 24$ $\mu$m tracing hot dust emission and H$\alpha$ tracing gas photoionized by young stars. The region appears compact at our resolution and shows an abundance of molecular gas and tracers of recent star formation (Table \ref{tab:jewel}) indicative of a young, intense star-forming knot. The Jewel shows up as a less-prominent but identifiable bright feature in the 3.6 $\mu m$ map. This may be the light from the associated stellar populations, or could reflect the contributions of 3.3 $\mu m$ PAH feature to the \textit{Spitzer} IRAC 1 band, adding evidence that the region is young with the stars still significant shrouded by dust \citep{2025ApJ...982...50W,2025ApJ...983..137R}.

Alongside the Jewel, the East Ring Pileup also contains a second significant star formation source we refer to as Region B. This is the second brightest source of both 24 $\mu$m and H$\alpha$ emission in the map, although it is more extended and has a lower surface density than the Jewel (Table \ref{tab:jewel}). 

The morphology of the East Ring Pileup is remarkably similar across the H$\alpha$ and 24$\mu$m maps in Figure 14, as well as the 6 GHz emission in Figure 4d of \citet{2019A&A...632A..12S}. This correspondence indicates that the emission at all three wavelengths traces the same underlying star-forming structures. The 6 GHz radio continuum is likely dominated by thermal free–free emission associated with ionized gas, consistent with its close spatial agreement with the H$\alpha$ emission, although a non-thermal contribution from cosmic ray electrons and magnetic fields may also be present. The similarity with the 24 $\mu$m emission further suggests that the ionized gas and warm dust are co-spatial and powered by the same young stellar population.

\begin{deluxetable}{l|cc} \label{tab:jewel}
\tablecaption{East Ring Pileup Properties}
\tablehead{
  \colhead{Property} & \colhead{Region A} & \colhead{Region B} 
}
\startdata
$\Sigma_{\rm mol}$  [$M_\odot pc^{-2}$]&268&174\\
$\sigma_v$ [km s$^{-1}$]&12.73 &10.93\\
$M_{mol}$ [$10^6M_\odot$]&24.0&32.2\\
$M_{*}$ [$10^7M_\odot$]&17.8&28.8\\
$I_{\mathrm{H}\alpha,\:peak}$ [$\rm 10^{-17}$ erg $\rm s^{-1}\:cm^{-2}$] &8.18&3.56\\
$\Sigma_{\mathrm{SFR}}$ [$M_\odot pc^{-2} Gyr^{-1}$]&46.5&50.0\\
$R_{\mathrm{eff}}$\tablenotemark{a} [pc]&169&243\\
$^{13}$CO/$^{12}$CO&0.160&0.142 \\
\enddata
\tablenotetext{a}{Derived from $R=\sqrt{A/\pi}$ where A is the area where $H_\alpha>3\sigma$.}
\end{deluxetable}

Compared to the broader GMC population in NGC 4565, the Jewel is a clear outlier. Its $\Sigma_{\rm mol} = 268\: {\rm M}_\odot \: {\rm pc}^{-2}$ lies well above the typical value for GMCs in NGC 4565. Region B is also above the average NGC 4565 surface density, though it is less dense than the Jewel.

When placed in a broader extragalactic context, the Jewel’s surface density exceeds the typical values seen in the Milky Way and M31, and is comparable to the upper envelope of the PHANGS GMC sample. 
This highlights the Jewel as a rare example of a very dense molecular complex within a relatively quiescent disk environment. 

Within the context of other well-known molecular clouds, the Jewel has a similar surface density to 30 Doradus in the Large Magellanic Cloud which has $\Sigma_{\rm mol}\sim 250\,M_\odot\,\mathrm{pc^{-2}}$ \citep{2011ApJS..197...16W}, although 30 Doradus has a much higher $\Sigma_{\rm SFR}$ \citep[][$\rm \Sigma_{SFR}\approx 1M_\odot\;yr^{-1}\: kpc^{-2}$ kpc]{2018Sci...359...69S} as a site of super star cluster formation. Among extragalactic analogs in quiescent disks, the Jewel is also comparable to the Headlight Cloud in NGC 628, which has a similarly high surface density \citep[$\Sigma_{\rm mol} \sim 192\,M_\odot,\mathrm{pc^{-2}}$;][]{2020A&A...634A.121H}. In contrast, the Jewel is denser than local molecular clouds like Orion A, and Orion B which have molecular gas surface densities of 80~$M_\odot\,\mathrm{pc^{-2}}$\citep{2008ApJ...680..428G}

On the more extreme end, both the Jewel and 30~Doradus remain far less dense than the Firecracker in the Antennae galaxies, a pre–super star cluster candidate with $\Sigma_{\rm mol}\gtrsim1800\ M_\odot\ \mathrm{pc^{-2}}$.  The Jewel is more comparable to the non-Firecracker clouds in the Antennae Overlap, which have average surface densities of $\Sigma_{\rm mol}\approx 441\ M_\odot\ \mathrm{pc^{-2}}$
\citep{2019ApJ...874..120F,2024ApJ...964..166K}. Thus, the Jewel is a particularly active region to be found in the disk of a main sequence galaxy, but less extreme than the regions found in rarer systems like major mergers. 

In many barred galaxies, star formation in rings is attributed to bar-driven gas inflows that funnel material inward along dust lanes and deposit it near resonance radii (e.g., \citealt{1992MNRAS.259..345A,2013ApJ...769..100S,2023MNRAS.523.2918S}). While NGC 4565 does host a stellar bar \citep{2019ApJ...872..106K}, we do not detect significant molecular gas interior to the ring that would indicate ongoing inflow feeding the structure. Additionally, the bar is oriented north–south while the Jewel is on the east part of the ring away from the bar ends. Upcoming JWST observations will allow us to probe the detailed ISM structure with $\sim$0.35" (20 pc) 7.7 $\rm\mu m$ imaging and localize the compact ionized gas with $\sim 0.1\arcsec$ (6 pc) Pa$\alpha$ observations.

\section{Summary}\label{sec:conc}

We present $0.94\arcsec \approx 55$~pc $^{12}$CO(2-1) and $^{13}$CO(2-1) observations of the highly inclined ($i\sim87.5^\circ$) galaxy NGC 4565. These data cover radii out to $\rm R_{gal} = \pm 17.5$ kpc and we use them to investigate the kpc- and cloud-scale vertical structure of the molecular gas, the link between environment and GMC properties, and the make-up of the ISM out to large $R_{\rm gal}$. Our key findings are:

\begin{enumerate}

\item $^{12}$CO emission is detected across the full extent of our map, out to $R_{\mathrm{gal}} \gtrsim 17.5$ kpc [\S \ref{sec:obs} and \S \ref{sec:radial_analysis}]. At our $\sim$55 pc, the emission resolves into GMC-scale structures at all radii [\S \ref{sec:gmcs}].

\end{enumerate}

We leverage the large radial extent and good sensitivity of our data to measure the large-scale structure of the molecular ISM in NGC 4565:

\begin{enumerate}
\setcounter{enumi}{1}

\item NGC 4565 has very little molecular gas inside its ring ($\rm R_{gal}\lesssim4~kpc$). This is followed by an $H_2$-dominated disk ($\rm 4 ~kpc\lesssim R_{gal} \lesssim 7 ~kpc$) and an HI-dominated outer disk ($\rm R_{gal}\gtrsim 7~kpc$). In terms of radial profiles, we note some similarity to M31 and the PHANGS--ALMA target NGC 2775 [\S \ref{sec:radial_analysis}].

\item The stellar, CO, and mid-IR radial profiles of NGC 4565 extend far out with long scale lengths ($l_{mol}\approx6.8$ kpc and $l_{*}\approx4.5$ kpc) reflecting the galaxy's high mass and large size. [\S \ref{sec:radial_analysis}]

\item The $^{13}$CO/$^{12}$CO line ratio is approximately flat ($\sim$0.086 $\pm$ 0.009) over $R_{\rm gal} = 5{-}13$~kpc, consistent with uniform optical depth and isotopologue abundance (or offsetting trends) across the disk. Over the whole radial range studied, our data are consistent with a gradient of $-0.036$~dex~kpc$^{-1}$ [\S \ref{sec:12to13}]
\end{enumerate}

Analyzing the detailed structure of the CO emission we find:

\begin{enumerate}
\setcounter{enumi}{5}

\item The $^{12}$CO emission is too clumpy for vertical profiles to be fit by smooth disk models. We instead use multi-Gaussian fits to position-velocity slices to characterize individual GMC-scale structures. This reveals a thin ($\sigma_y \approx 33.6\pm0.7$ pc, FWHM $\approx 79.1\pm1.6$~pc) molecular disk. We observe minimal vertical flaring with increasing radius, but emphasize that our method captures the vertical extent of the individual features (i.e., GMCs) and not the cloud-to-cloud vertical scatter [\S \ref{sec:methods2} and \S \ref{sec:hmol}].

\item Giant molecular clouds in NGC 4565 have sizes, velocity dispersions, surface densities, and virial parameters that generally align with the distributions of physical properties in less inclined galaxies. We observe a decline in $\sigma_v$ with increasing galactocentric radius consistent with the sense of results for other galaxies but extending to very large radius in NGC 4565 [\S \ref{sec:gmccomp}].

\item The properties of GMCs in NGC 4565 show a similar correlation with their larger-scale environment (we emphasize $\Sigma_{\rm mol, kpc}$) to PHANGS--ALMA galaxies, as do GMCs in archival M31 data and the Milky Way. The strongest correlations relate cloud-scale $\Sigma_{\rm mol}$ and $\sigma_v$ to $\Sigma_{\rm mol, kpc}$ [\S \ref{sec:env}].


\item  Based on a \texttt{CPROPS} object-finding analysis, clouds are preferentially aligned with the major axis of the galaxy with moderate axis ratios, $\eta_R^{-1} \approx 1.5$. Comparing cloud properties among galaxies with a range of inclinations shows that statistical population-level inclination effects appear small for $R$ and $\alpha_{vir}$ and moderate for $\sigma_v$. [\S \ref{sec:incl}]

\item The ``East Ring Pileup'' marks a prominent departure from axisymmetry. This includes a compact region that we dub the ``Jewel,'' which is bright at many wavelengths. The Jewel is comparable in density to Local Group starburst regions like 30 Doradus, though not as dense as the star-forming regions found in even more extreme environments. [\S\ref{sec:obs} and \S\ref{sec:jewel}]

\end{enumerate}

The full data CO data cubes and cloud property tables will be made publicly available via CADC.

\begin{acknowledgments}
We thank the anonymous referee for a detailed, timely, and constructive report.
This paper makes use of the following ALMA data: ADS/JAO.ALMA \#2018.1.01050.S. ALMA is a partnership of ESO (representing its member states), NSF (USA) and NINS (Japan), together with NRC (Canada), MOST and ASIAA (Taiwan), and KASI (Republic of Korea), in cooperation with the Republic of Chile. The Joint ALMA Observatory is operated by ESO, AUI/NRAO and NAOJ. The National Radio Astronomy Observatory is a facility of the National Science Foundation operated under cooperative agreement by Associated Universities, Inc. G.K was supported by NSF Graduate
Research Fellowship (DGE-2146755). ER acknowledges the support of the Natural Sciences and Engineering Research Council of Canada (NSERC), funding reference number RGPIN-2022-03499.
\end{acknowledgments}

\appendix

\section{Height Above the Midplane}\label{sec:yoff}

The less-than-edge-on orientation of NGC 4565, with $i = 87.5^\circ$ instead of $90^\circ$, results in ambiguity interpreting distances along the $y$ axis. A particular point of confusion is that the $y$-offset of the fitted Gaussians in our analysis (see Figures \ref{fig:vertprof} and \ref{fig:bins}) can either correspond to a physical height above the galactic midplane, or the foreshortened in-plane position along the minor axis. This ambiguity prevents us from assessing the vertical distribution of clouds.

In order to break this degeneracy, we employ the velocity information to make an independent measurement of the galactocentric radius of each component, $R_{\rm vel}$. This utilizes the $x$-offset and velocity to infer a galactocentric radius based on the galaxy's rotation curve (Section \ref{sec:radcomp}). We contrast this with the geometric radius measurement, $R_{dpj}$, which uses the $x$ coordinate as a major axis offset and the $y$ coordinate as the foreshortened in-plane position along the minor axis such that 

\begin{equation}
    R_{dpj} = \sqrt{x^2+(\frac{y}{\cos i})^2}.
\end{equation}

Assuming that the velocity-based radius captures the true galactocentric radius of a cloud, the slight discrepancies between $R_{\mathrm{vel}}$ and $R_{\mathrm{dpj}}$ seen in Figure \ref{fig:radcomp} could be driven by the vertical height of that cloud above the midplane. This vertical height will contribute to $R_{dpj}$ but not $R_{\rm vel}$.

To approximate the vertical distance of a given point above or below the midplane, we project the radial difference between these two estimates back into the vertical direction. Using the $y$-contribution to each radius measurement, we have:

\begin{equation}
    y^2 = (R_{\rm vel}^2-x^2)\cos^2 i,
\end{equation}

\begin{equation}
    \text{and }(y+z)^2 = \sqrt{(R_{dpj}^2-x^2)\cos^2i}.
\end{equation}

This yields:

\begin{equation}
\label{eq:height}
    \text{Height above midplane: } z= \cos i \left ( \sqrt{R_{dpj}^2-x^2} - \sqrt{R_{\rm vel}^2-x^2} \right ).
\end{equation}

This relation provides an upper limit on the physical height, since any departure from purely circular motion or local non-circular streaming could also contribute to deviations in $R_{\rm vel}$. However it is useful for placing stricter constraints on the height of structures above the midplane. 

\begin{figure}
    \centering
    \includegraphics[width=0.8\linewidth]{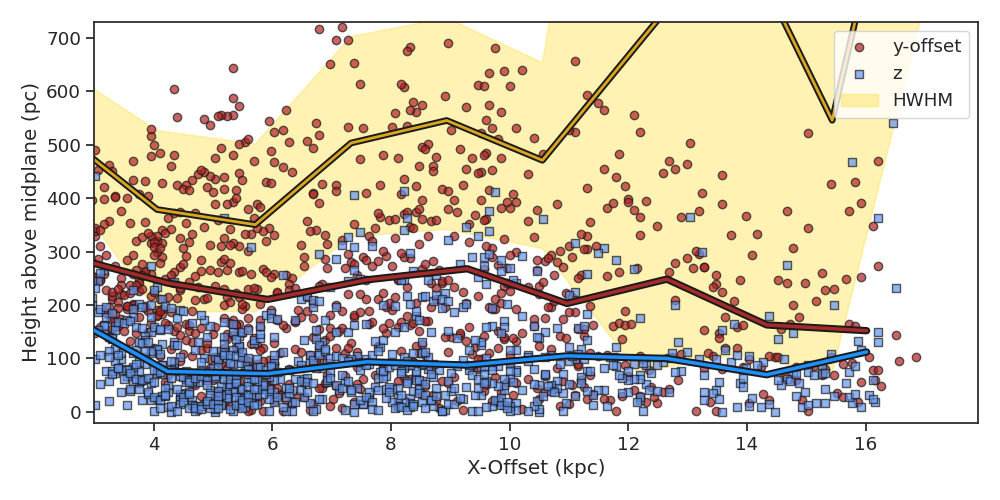}
    \caption{Height above the midplane of individual clouds. Red points show the results (incorrectly) attributing all of the $y$ coordinate to vertical location. Blue points show results from contrasting position+velocity measurements with the ($x$, $y$) location (Equation \ref{eq:height}). The solid lines represent the median height in bins of $x$ for each approach. The yellow shading shows the 16-84\% range of the HWHM of the full vertical profile (e.g., Figure \ref{fig:vertprof}) at each $x$-offset.}
    \label{fig:yoff}
\end{figure}

In Figure \ref{fig:yoff}, we compare the $y$-offset from the vertical profiles to the estimated height above the midplane, $z$. The two quantities show similar trends across the disk, remaining relatively flat as a function of $x$-offset. However, the inferred vertical height $z$ is consistently much smaller in magnitude than the total $y$-offsets, with median $z$ of 97~pc and median $y$ of 244 pc. This indicates that much of the apparent vertical displacement in the profile is due to in-plane projection effects rather than vertical departures from the midplane. The lower $z$ values are also more on the scale of Milky Way midplane displacement measurements, which are typically less than 400 pc, and much lower at low Galactocentric radius \citep[see][where  the Milky Way shows significant flaring at larger radii]{2015ARA&A..53..583H}. 

\section{GMC Properties from \texttt{CPROPS}\label{sec:cprops}}

In addition to our position-velocity diagram fitting, we identify and characterize clouds in the CO(2-1) data cube using the python implementation of \texttt{CPROPS} \citep{2006PASP..118..590R}. The current implementation of \texttt{CPROPS} \cite{2021MNRAS.502.1218R} masks the data cube, identifies maxima in position–position–velocity space, and then segments the cube using a watershed algorithm. Then the software measures cloud properties such as size, line width, and luminosity based on moment calculations. \texttt{CPROPS} has been used in many extragalactic GMC studies, including the PHANGS--ALMA and M31 datasets used in Section \ref{sec:gmccomp}, and this exercise allows us to compare clouds from NGC 4565 to those other systems using matched methodology.

\begin{figure*}
    \centering
    \includegraphics[width=0.9\textwidth]{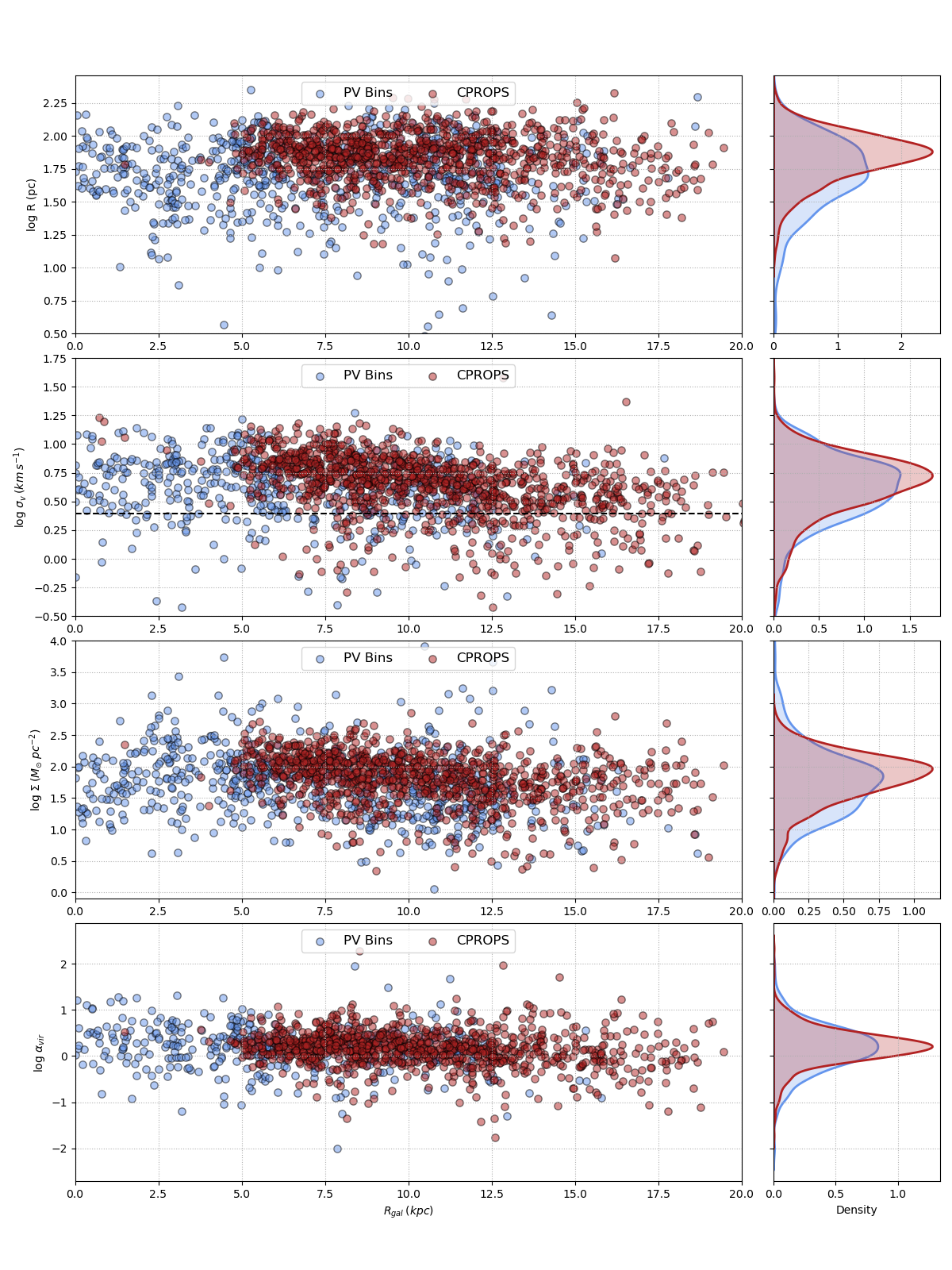}
    \caption{Distributions of GMC properties as a function of galactocentric radius comparing the multi-gaussian decomposition method to the \texttt{CPROPS} algorithm.}
    \label{fig:cprops_comp}
\end{figure*}

In Figure~\ref{fig:cprops_comp}, we compare the distributions of GMC size, velocity dispersion, density, and virial parameter as a function of galactocentric radius derived using \texttt{CPROPS} and our multi-Gaussian decomposition method. The two methods show similar distributions and trends with radius, but there are some differences. Among the most noticeable differences across the different properties are that \texttt{CPROPS} does not identify many GMCs at small galactocentric radii and the overall property distributions are narrower than those of GMCs identified through fitting multiple Gaussians to our vertical profiles. This leads to the radial trend of declining $\Sigma_{\rm mol}$ appearing even clearer in the \texttt{CPROPS} measurements compared to our fiducial decomposition. Additionally, while the average GMC properties tend to align for both methods, \texttt{CPROPS} does tend to identify larger structures with average radii 1.4 times larger. This is due to the \texttt{CPROPS} algorithm building a zone of avoidance around identified peaks.

Another important difference between the two methods is that our approach in the main text fits only the minor axis $y$ size of each feature. \texttt{CPROPS} estimates the major and minor axis of the emission in on the sky and also returns a position angle and aspect ratio. As shown in Figure~\ref{fig:cprops_posang}, the majority of \texttt{CPROPS}-identified GMCs have position angles clustered around the major axis of the galaxy, although this is not universal --- we find clouds with a wide range of position angles, indicating that some structures are significantly misaligned with the galactic plane.

We also plot the ratio of the major to minor axes (i.e., the cloud aspect ratio) as a function of position angle. At first glance, the distribution of individual clouds suggests that GMCs with position angles close to the galaxy’s major axis tend to have larger aspect ratios, potentially indicating greater elongation along the disk. However, when the data are binned by position angle, the average aspect ratio remains relatively flat across all angles. This indicates that there are more GMCs with larger aspect ratios aligned with the disk, not by nature of their alignment, but because there are simply more GMCs aligned with the disk in general and thus are more reflective of the true scatter in aspect ratio.




\bibliography{bibliography}{}
\bibliographystyle{aasjournal}

\end{document}